\documentclass[letterpaper]{JHEP3}

\usepackage{amsmath}
\usepackage{amsfonts}
\usepackage{graphicx}

\def\bea{\begin{eqnarray}}
\def\eea{\end{eqnarray}}
\def\nn{\nonumber}

\def\endignore{}
\def\ignore #1\endignore{} 

\def\bd{\begin{displaymath}}
\def\ed{\end{diplaymath}}

\def\cW{{\cal W}}

\def\cN{{\cal N}}
\def\cF{{\cal F}}

\def\ba{\begin{eqnarray}}
\def\ea{\end{eqnarray}}
\def\be{\begin{equation}}
\def\ee{\end{equation}}

\newcommand{\N}{\mathcal{N}}

\newcommand{\del}{\partial}
\newcommand{\bi}{\bar{\imath}}
\newcommand{\bj}{\bar{\jmath}}
\newcommand{\comment}[1]{}

\newcommand{\beq}{\begin{equation} }
\newcommand{\eeq}{\end{equation}}

\newcommand{\roughly}[1]{\raise.3ex\hbox{$#1$\kern-.75em\lower1ex\hbox{$\sim$}}}

\newcommand{\bt}{\begin{tabular}}
\newcommand{\et}{\end{tabular}}
\newcommand{\bc}{\begin{center}}
\newcommand{\ec}{\end{center}}

\newcommand{\cof}{{\rm cof}}
\newcommand{\Weff}{W^{\rm (eff)}}
\newcommand{\Weffb}{\bar W^{\rm (eff)}}

\newcommand{\Ox}{\Omega}

\newcommand{\cB}{\mathcal{B}}

\newcommand{\cJ}{\mathcal{J}}

\newcommand{\cM}{\mathcal M}

\newcommand{\ib}{{\bar\imath }}
\newcommand{\jb}{{\bar\jmath }}

\newcommand{\IM}{\textrm{Im} \,}
\newcommand{\RE}{\textrm{Re} \,}

\newcommand{\cref}{{\bf [check ref]}}

\newcommand{\eff}{\rm (eff)}

\newcommand{\hi}{\hat{\imath}}
\newcommand{\hj}{\hat{\jmath}}
\newcommand{\NJ}{{\cal N}_J}
\newcommand{\NO}{{\cal N}_\Omega}

\title{No-scale supersymmetry breaking vacua and soft terms with torsion}

\author{P.G. C\'amara,$^1$ M. Gra\~{n}a,$^2$
\\

$^1$ Centre de Physique Th\'eorique, UMR du CNRS 7644,

Ecole Polytechnique,

91128 Palaiseau, France.

E-mail: \textup{\texttt{pablo.camara@cpht.polytechnique.fr}}\\

$^2$ Service de Physique Th\'eorique,

CEA/Saclay,

91191 Gif-sur-Yvette Cedex, France.

E-mail: \textup{\texttt{mariana.grana@cea.fr}}\\ }
\date{}

\abstract {We analyze the conditions to have no-scale
supersymmetry breaking solutions of type IIA and IIB supergravity
compactified on manifolds of $SU(3)$-structure. The supersymmetry
is spontaneously broken by the intrinsic torsion of the internal
space. For type IIB orientifolds with O9 and O5-planes the mass of
the gravitino is governed by the torsion class $\mathcal{W}_1$,
and the breaking is mediated through F-terms associated to descendants of the
original $\mathcal{N}=2$ hypermultiplets. For type IIA
orientifolds with O6-planes we find two families of solutions,
depending on whether the breaking is mediated exclusively by hypermultiplets
or by a mixture of hypermultiplets and vector multiplets, the
latter case corresponding to a class of Scherk-Schwarz
compactifications not dual to any geometric IIB setup. We compute
the geometrically induced $\mu$-terms for D5, D6 and D9-branes on
twisted tori, and discuss the patterns of soft-terms which arise
for pure moduli mediation in each type of breaking. As for D3 and
D7-branes in presence of 3-form fluxes, the effective scalar
potential turns out to possess interesting phenomenological properties.}

\preprint{CPHT-RR153.1007\\  SPhT-T07/135 \\ arXiv:0710.4577}

\begin{document}

\section{Introduction} \label{intro}

Compactifications with fluxes (see \cite{RF} for reviews) have
been intensively studied in the past few years for their potential
phenomenological applications. They provide us with powerful tools
for finding stable or metastable vacua within string theory. The
topological requirement for a reduction to a four-dimensional low
energy supersymmetric effective theory is that the internal
manifold should allow a nowhere vanishing spinor. Six-dimensional
manifolds admitting nowhere vanishing spinors have structure group
reduced to SU(3) or subgroups of it \cite{CS,Gstructures}. On
manifolds of G-structure, there is always a torsionful connection
under which the G-invariant spinor is covariantly constant.
Torsion leads to non-integrability of the G-structure, and can be
thought as another NSNS flux of the theory (and thus is sometimes
called ``(geo)metric flux"). In some situations it can be dual to
NSNS 3-form flux $H$.

The effective $\N=1$ theory for reductions on (orientifolds of)
SU(3)-structure manifolds
\cite{halfflat,GuMi,HoPa,GLW,iman,KPM,GLW2,Am2} is defined by a
K\"ahler potential and a superpotential. The former describes the
space of moduli consisting of variations of the RR potentials and
deformations of B-field and metric in the class of SU(3) structure
manifolds. The superpotential involves all the fluxes, RR and NSNS
($H$ and torsion). Supersymmetric vacua can be found either by
varying the action \cite{DKP,VZ,Dw,cfi,robbins1,robbins2}, or directly by solving
the six-dimensional internal first order differential equations
for supersymmetric vacua \cite{gmpt2}.  The two procedures have
been shown to be equivalent  \cite{4to10}.

After the huge progress achieved in understanding the conditions
for supersymmetric vacua and finding examples, the path continues
by exploring the mechanisms of supersymmetry breaking, and finding
stable or metastable non-supersymmetric vacua. Spontaneous
breaking of supersymmetry can be generated dynamically, or be
present already at tree level. Long-lived metastable vacua with
dynamically broken supersymmetry were found in SQCD  \cite{ISS},
and their stringy realizations proposed in
\cite{smeta0,smeta1,smeta2}. On the other hand, supergravity vacua
with supersymmetry broken at tree level have been mostly
considered in the framework of type IIB supergravity compactified
in Calabi-Yau orientifolds with O3-planes and a imaginary
self-dual combination of NSNS and RR 3-form fluxes \cite{GP}.
There, the amount of supersymmetry breaking is a tunable parameter
which can be set to zero, and the solutions can be seen as
``marginal'' deformations of $\mathcal{N}=1$ vacua. From the
four-dimensional point of view, the sector that breaks
supersymmetry only involves moduli descending from $\mathcal{N}=2$
hypermultiplets (which in this case correspond to deformations of
the K\"ahler form). The cosmological constant vanishes at tree
level, and the supersymmetry breaking vacua are of  no-scale
type \cite{noscale}. At the quantum level, however,
non-perturbative effects may lift the remaining flat directions
and restore supersymmetry in an AdS vacuum. This AdS point has
been the basic building block in many of the recent attempts to
address the problem of supersymmetry breaking within
compactifications of string theory, and also for building up
models with de-Sitter minima \cite{KKLT}.

When D-branes are present, spontaneous breaking of supersymmetry
in the bulk is communicated to the brane sector by the moduli in
the closed string sector (neutral matter), and manifests in the
open string sector (charged matter) as soft-breaking of
supersymmetry \cite{IL,KL,BIM}. Soft-supersymmetry breaking terms
for D-branes on Calabi-Yau manifolds or tori  in the presence of
supersymmetry breaking fluxes have been obtained in
\cite{soft1,soft2,hepth0311241,GGJL,0406092,hepth0408036,0410074,fontibanez,0501139}.
These papers show that no-scale spontaneous breaking of
supersymmetry by imaginary self-dual 3-form fluxes is not
communicated to  D3-branes. However, for  branes wrapping internal
dimensions, richer patterns of soft-terms arise.

Motivated by all these results, in this paper we look for classes
of no-scale supersymmetry breaking vacua involving orientifolds
(O5/O9 and O6) of manifolds of SU(3) structure. Supersymmetry is
broken spontaneously by fluxes and torsion. The supersymmetry
breaking vacua we discuss are basically divided into two types,
one where the supersymmetry breaking sector lies entirely in
descendants of ${\mathcal N}=2$ hypermultiplets, and another one where the
breaking sector involves the two type of moduli, those descending
from hypermultiplets and those from vector multiplets. The former
are believed to be T-dualizable to O3 setups of the class in
\cite{GP}, while the latter, which includes Scherk-Schwarz
mechanism \cite{schsch}, seem to have non-geometric O3 duals. We
illustrate each class with examples in toroidal models. Similar
toroidal no-scale vacua were obtained from the four-dimensional
low-energy action in \cite{DKP,cfi}, whereas some previous work
on no-scale supersymmetry breaking in supergravity compactifications
was carried out in \cite{andrianopoli}.

In a similar fashion than $H$ and RR fluxes, metric fluxes induce
$\mu$-terms as well as soft-supersymmetry breaking terms on
D-branes. In the article we also study the effect of torsion on
D5, D9 and D6-branes in toroidal models. Using the brane
superpotentials of \cite{mar1},  we find the torsion induced
$\mu$-terms. Finally, we analyze the soft-supersymmetry breaking
patterns for the classes of supersymmetry breaking vacua discussed
previously, for pure moduli mediation.

The paper is organized as follows. In section \ref{gen} we show
the basic features of compactifications on manifolds of SU(3)
structure. We use concepts of generalized complex geometry
\cite{Hitchin,gualtieri}, which we find best adapted to describe
the bulk and brane physics.  In section \ref{bulkIIB} we discuss
supersymmetry breaking no-scale vacua in IIB, illustrate with
examples, and find the induced $\mu$-terms on D9 and D5-branes. In
section \ref{bulkIIA} we analyze the IIA counterparts. In section
\ref{soft} we study the soft-supersymmetry breaking patterns for
the two classes of supersymmetry breaking mechanisms discussed in
sections \ref{noscaleIIB} and \ref{noscaleIIA}. Section
\ref{concl} contains a summary and conclusions. Appendices
\ref{conv}, \ref{su3decomp} and \ref{torsiontori} present some of
the conventions used, as well as some technical details needed in
the text.

\section{Compactifications with SU(3) structure and D-branes / orientifold planes} \label{gen}

In this section we review the necessary features about Minkowski
compactifications on manifolds of SU(3) structure, and
supersymmetric D-branes on them.

\subsection{Type II supergravities on SU(3) structure manifolds: bulk} \label{genbulk}

 We study warped compactifications on manifolds of SU(3) structure, i.e. the ten-dimensional metric
is given by \beq \label{metric} ds^2= e^{2A} \eta_{\mu \nu} dx^\mu
dx^\nu + ds^2_{\cM_6} \, , \quad \mu=0,1,2,3 \eeq where $e^{2A}$
is the warp factor. The internal manifold ${\cM_6}$ has SU(3)
structure \cite{CS,Gstructures}, which we define in the following
subsection, and use all throughout the text.

\subsubsection{SU(3) structure definitions}
\label{su3def}

On a manifold of SU(3) structure there is a globally defined SU(3)
invariant non-degenerate 2-form $J$, and a holomorphic 3-form $\Omega$
satisfying \beq \label{comp} J \wedge \Omega = 0 \ , \qquad J
\wedge J \wedge J = -i \, \frac34  \frac{\NJ}{\NO} \, \Omega
\wedge \bar \Omega\ , \eeq for some constants $\NJ$, $\NO$. The
former conditions say that $J$ is a (1,1)-form in the complex
structure defined by $\Omega$. The constants  $\NJ$, $\NO$ define
the normalization of $J$ and $\Omega$, in the following sense
\footnote{A usual convention is to take the ratio $\NJ / \NO=1$,
but here we find it more convenient to leave it unfixed.},
\begin{equation} \label{NJNO}
\mathcal{N}_J\equiv \frac{1}{6}\int J\wedge J\wedge J \quad , \quad \mathcal{N}_{\Omega}\equiv \frac{1}{8i}\int\Omega\wedge \overline{\Omega}\ .
\end{equation}
In this parametrization, $\mathcal{N}_J$ corresponds to the volume
of the manifold  whereas $\frac16J^3$ is the volume form. A very
important fact is that for manifolds with just SU(3) structure,
there are no globally defined 1-forms.

$J$ defines a symplectic structure $J_{mn}$ (a skew-symmetric map from $T \times
T$ to $\mathbb R$ with inverse $J^{-1}$) and $\Omega$ a complex
structure $I^m{}_n$ (a map from $T$ to itself that squares to -1).
Provided (\ref{comp}) is satisfied, both structures intersect on a
SU(3) structure.
 The complex structure $I$ can be read off
from the local decomposition of $\Omega$ in terms of holomorphic
1-forms $z^i$, namely we can write locally $\Omega= \frac{1}{6}
\epsilon_{ijk} z^i \wedge z^j \wedge z^k$. The dual vectors to
$z^i$, that we call $\del_{z^i}$, form a basis for holomorphic
vectors $v=v^i \del_{z^i}=v^m \del_{m}$, where $\del_m$ is a basis
of real coordinates. The complex structure should be such that a
vector constructed in this way is holomorphic, namely $I^m{}_n v^n
= i v^m$.

$J$ and $I$ (or equivalently $\Omega$) define a metric, given by
\beq g_{mn}=J_{mp} \, I^p{}_n \ , \eeq which is automatically
symmetric if the first condition in (\ref{comp}) is satisfied. The
SU(3) structure can be given alternatively by the metric and a
globally defined SU(3) invariant spinor $\eta$. Then, $J$ and
$\Omega$ can be constructed as bilinears of the spinor, as we show
in (\ref{bilinears}).

If the symplectic and holomorphic forms are closed, $dJ=0,$
$d\Omega=0$, the corresponding structures are integrable. For the
case of $\Omega$, this implies that there are local functions
$f^i$ such that the 1-forms $z^i=df^i$ (i.e., the equation $z=df$
is integrable). An analogous statement can be made with integrable
symplectic structures. In such case, the manifold has SU(3)
holonomy. On a generic SU(3) structure, none of the structures is
integrable, and therefore $dJ$ and $d\Omega$ are not zero. The
3-form $dJ$ and 4-form $d\Omega$ can be decomposed in SU(3)
representations, and the corresponding components are the torsion
classes, defined as \cite{CS}
\begin{align} \label{su3torsions}
dJ&=\frac{3}{2}\frac{\mathcal{N}_J}{\mathcal{N}_{\Omega}}\textrm{Im}(\mathcal{W}_1\overline{\Omega})+\mathcal{W}_4\wedge J+\mathcal{W}_3\ , \\
d\Omega&=\mathcal{W}_1J\wedge J+\mathcal{W}_2\wedge
J+\overline{\mathcal{W}}_5\wedge \Omega \ ,
\end{align}
where $\cW_1$ is a complex scalar, $\cW_2$ a complex primitive
(1,1) form, $\cW_3$ a real primitive $(2,1) + (1,2)$ form and
$\cW_4$ and $\cW_5$ real vectors ($\cW_5$ is actually a complex
(1,0)-form, which has the same degrees of freedom).

\subsubsection{SU(3) structures and generalized complex geometry}

Generalized complex geometry \cite{Hitchin,gualtieri} is a
suitable framework for describing IIA and IIB on the same footing.
We will give here just a very minimal review of it containing the
basic definitions we will use. More extensive reviews in the
context of flux compactifications can be found for example in
\cite{GLW,LMT,GMPT1,mar2,scan}.

Complex and symplectic structures can actually be defined by a
single type of structure: a generalized complex structure
\cite{Hitchin,gualtieri}. Generalized complex structures are
defined in an analogous way as standard complex structures, i.e.
as maps from a bundle to itself that square to $-1$. The bundle in
question is however extended (or generalized) to the sum of the
tangent plus cotangent bundles of the manifold, $T_{\cM} \oplus
T^*_{\cM}$. From $J$ and $I$ we can build the following
generalized complex structures, \beq \label{gencompl} {\cal J}_-=
\left( \begin{array}{cc} I & 0 \\  0 & -I^T  \end{array} \right) \
,\qquad  {\cal J}_+= \left( \begin{array}{cc} 0 & J^{-1} \\ -J & 0
\end{array} \right) \ , \eeq where the meaning of the subscripts
plus and minus will become clear later.

There is a one to one map between generalized complex structures
and  $O(6,6)$ pure spinors (i.e., spinors annihilated by half of
the Clifford(6,6) gamma matrices). Given a pure spinor $\Phi$, its
corresponding generalized complex structure ${\cal J}_\Phi$  is
such that the $+i$ eigenbundle of ${\cal J}_\Phi$ is the
annihilator of the pure spinor. In addition, $O(6,6)$ spinors are
isomorphic to sums of forms \footnote{The isomorphism is
$\gamma^{m_1 ... m_k} \cong e^{m_1} \wedge ... \wedge e^{m_k}$.}.
Positive (negative) chirality spinors are associated to even (odd) forms.

The $O(6,6)$ pure spinors corresponding to (\ref{gencompl}) are
\beq
 \quad \Phi_-= 8 e^{i \theta_-} \, \eta_+ \eta_-^{\dagger} = - i e^{i \theta_-} \left( \frac{{\cal N}_J}{\NO} \right)^{1/2} \, \Omega  \ , \qquad  \Phi_+=8 e^{i \theta_+}  \, \eta_+ \eta_+^{\dagger} =  e^{i \theta_+} e^{-iJ} \ , \label{SU3}
\eeq
where $\eta$ is the $O(6)$ spinor
defining the SU(3) structure. In order to get the forms, we have
used the Fierz identity (\ref{fierz}) and the bilinears in
(\ref{bilinears}).
Notice that $\Phi_-$ ($\Phi_+$) contains only odd (even) forms, as
it should be from their chiralities. This explains the use of plus
and minus in (\ref{gencompl}). Moreover, the one to one
correspondence between generalized complex structures and pure
spinors define the latter up to some overall complex number,
that we chose to be $8\textrm{exp}(i\theta_\pm)$.
These phases will be important later, and are fixed
by the orientifold projection.

The action of the B-field on the pure spinors can be encoded in
the ``B-transform'' of the spinor, $e^{-B} \Phi$, where $e^{-B}= 1 - B \wedge +\ (1/2) B
\wedge B \wedge + \ldots$.   It is not hard to show that if $\Phi$ is a pure spinor, then
$e^{-B} \Phi$ is also pure. This allow us to work in terms of some
new spinors $\tilde{\Phi}_\pm\equiv e^{-B}\Phi_\pm$ which define
not only the metric, but also the $B$ field. However, if $\Phi$ is closed,
$\tilde \Phi$ is generically not closed, as it contains $dB$. It is useful to define
the twisted exterior derivative $d_H=d-H\wedge$, where $H$ may
also contain a possible background flux $\bar H$, such that
\begin{equation}
d\tilde \Phi\equiv d(e^{-B}\Phi)=e^{-B}d_H\Phi\ .
\end{equation}
 If $\tilde{\Phi}$ is
integrable with respect to $d$, so is $\Phi$ with respect to
$d_H$. Note that the B-field action does not modify $\Phi_-$ since
for an SU(3) structure $B$ has to be (1,1), and therefore $B
\wedge \Omega = 0$. We leave it nevertheless to include the action
of $H$ on the exterior derivative.

The B-transform of a given pure spinor is associated to the B-transform of its generalized complex structure, given by
\beq
  \cJ^B={\cal B} \cJ \cB^{-1} \ , \qquad  \cB= \left( \begin{array}{cc} 1 & 0 \\ -B & 1 \end{array} \right) \ .
  \eeq
Since $B$ is a (1,1)-form, the matrix $BI$ is symmetric
and therefore $\cJ^{B}_-=\cJ_-$. The
generalized complex structure $\cJ_+$ on the contrary is modified
to \beq \label{gensympl} {\cal J}^B_{+}= \left(
\begin{array}{cc} J^{-1} B & J^{-1} \\ -(J+B J^{-1} B) & -B J^{-1}
\end{array} \right) \ . \eeq In what follows we will mainly work
with the polyforms (\ref{SU3}), although for some particular
purposes, such as the moduli definitions, it will be more
convenient instead to make use of $\tilde{\Phi}_\pm$.

\subsubsection{Orientifold projection and ${\cal N}=1$ vacua}

No-go theorems for Minkowski compactifications imply that whenever
fluxes are turned on, we need sources of negative charge and
tension if the internal manifold is compact. We therefore study
compactifications on orientifolds of manifolds of SU(3) structure,
concentrating on O5/O9 compactifications of type IIB, and O6
compactifications of type IIA. The orientifold projection is the
selection of even states under the combined action of the
world-sheet parity $\Omega_{\textrm{P}}$ and an involution
$\sigma$ (for consistency, an additional factor of $(-1)^{F_L}$ is
needed for O3/O7 and O6 projections). The
 involution $\sigma$ should be holomorphic in IIB ($\sigma I=I$) and antiholomorphic in IIA
 ($\sigma I=-I$). This leads to the following action on $J$ and $\Omega$ \footnote{The O6 projection allows $\sigma \Omega=e^{i \theta} \bar \Omega$. Here we are fixing $\theta=0$.} \cite{hepth0202208,hepth0303135,hepth0401137}
 \begin{align}
 \textrm{O6:} \quad &  \sigma J=-J \ , & &\sigma\Omega=\bar \Omega \ , \nn \\
 \textrm{O5/O9:} \quad & \sigma J=J \ ,  & &\sigma \Omega=\Omega \ .
 \end{align}
We can think of this as an action on the pure spinors as follows:
the world-sheet parity operator exchanges left and right movers,
which on the bispinors in (\ref{SU3}) has the effect of a
transposition\footnote{\label{SU3phases}$\Phi_\pm$ in (\ref{SU3})
should be thought as $\eta_{L+} \eta_{R \pm}^\dagger$. On a
manifold of SU(3) structure there is only one globally defined
spinor $\eta$, and therefore we have, up to overall normalization
that we fix to $1$, $\eta_{L+}=e^{i \theta_L} \eta_+ \ ,
\eta_{R+}=e^{i \theta_R} \eta_+$. The phases in (\ref{SU3}) are
$\theta_\pm \equiv \theta_L \mp \theta_R$ }. On the forms
associated to the bispinors, this transposition amounts to
conjugation plus some signs, which are conveniently encoded in the
operator \beq \label{lambda} \lambda (A)= \sum_n (-1)^{[(n+1)/2]}
A_n\ , \eeq where the subindex $n$ denotes the degree of the form
and $[\ldots]$ is the integer part. The action of $\sigma$ on the
forms (\ref{SU3}) is therefore \cite{iman},
\begin{align}
\textrm{O6:} \quad &\sigma \Phi_+=  \lambda (\Phi_+) \ ,  &  & \sigma\Phi_-=\lambda(\bar \Phi_-) \ , \nn \\
\textrm{O5/O9:} \quad & \sigma \Phi_+= -\lambda(\bar \Phi_+) \ , &
&\sigma \Phi_-= \lambda(\Phi_-)\ .
\end{align}
The phases $\theta_\pm$ in (\ref{SU3}) are then fixed to
$\theta_+=0$, $\theta_-=\pi/2$ for O6, and $\theta_\pm=\pi/2$,
for O5/O9.

The equations for $\cN =1$ Minkowski vacua in terms of $\Phi_\pm$
are \cite{gmpt2}, \bea \label{N=1vacua} d_H (e^{3A-\phi}\Phi_1)&=&
0 \, , \nn \\
d_H (e^{3A-\phi}\Phi_2)&=&-
 \label{nonint} e^{3A-\phi} dA\wedge\bar\Phi_2
- e^{3A} *\lambda(F)\
\eea
where
\begin{align}
\textrm{IIA:} \quad & \Phi_1= \Phi_+ \ ,  &  &\Phi_2= \Phi_- \nn \\
\textrm{IIB:} \quad &  \Phi_1= \Phi_- \ ,  &  &\Phi_2= \Phi_+  \ .
\end{align}
The RR form  $F$ is a purely internal form related to the total
ten--dimensional RR field strength,
\begin{equation}
    \label{F10F6}
F^{(10)} = F+ \mathrm{vol}_4\wedge \lambda(*F)\ , \qquad
F=\begin{cases}
F_0+F_2+F_4+F_6   & ({\rm IIA})\\
F_1+F_3+F_5 & ({\rm IIB})
\end{cases}
\end{equation}
where $\lambda$ is defined in (\ref{lambda}), $*$ is the
six-dimensional Hodge dual and the RR field strengths
\begin{equation}
F_{n}=dC_{n-1}-H\wedge C_{n-3}+e^B\bar F\ \label{fstr}
\end{equation}
satisfy the Bianchi identity $d(e^{-B}F)=0$ in absence
of localized sources.

Equations (\ref{N=1vacua}) tell us that $\cN=1$ Minkowski
vacua require one closed pure spinor, whose parity is equal to
that of the RR fluxes. The latter act as an obstruction for
integrability of the real part of the other pure spinor.

\subsubsection{K\"ahler potential, superpotential and no scale vacua}
\label{KWgen}

The moduli for $\cN=1$ compactifications arrange in chiral
multiplets. The orientifold projection splits the  $\cN=2$ vector
multiplets into $\cN=1$ vector and $\N=1$ chiral multiplets. The
latter contain the scalars parameterizing variations of $\Phi_1$.
The $\cN=2$ hypermultiplets parameterize variations of $\Phi_2$
(plus the dilaton and axion $B_{\mu \nu}$), paired with the axions
from RR scalars. The orientifold projection keeps only the
variations in the real part of $\Phi_2$, and combines them with
the surviving RR scalars, in the poly-form \cite{iman}
\beq \label{Pi} \Pi = {\cal C} + i \,
e^{-\phi} \, \RE \Phi_2 \ , \eeq with ${\cal C}$ the sum of RR
potentials (which have the same chirality as the form $\Phi_2$).

The $\cN=1$ K\"ahler potential is \cite{iman} \beq \label{Kgen}
K=- \textrm{log} \left[ -i \int \langle \Phi_1, \bar \Phi_1
\rangle \right]  - 2 \, \textrm{log} \left[ -i \int \langle \Phi_2,
\bar \Phi_2 \rangle \right] -2\textrm{log} \left(e^{-2\phi}\right)\eeq
where the Mukai pairing is defined as \beq \label{Mukai} \langle A
, B \rangle = (-1)^{[(n+1)/2]} A_n \wedge B_{6-n} = [ \lambda(A)
\wedge B ]_6\ . \eeq
$\Phi_1$ and $\Pi$ (or more precisely their B-transforms,
$\tilde{\Phi}_1$ and $\tilde{\Pi}$) should be expanded in a basis
of even or odd forms under the orientifold involution, according
to the case. The moduli in $\tilde \Phi_1$ descend directly from their
$\cN=2$ counterparts (but only those corresponding to
forms with the appropriate parity survive). As for $\tilde \Pi$, in order
to get a K\"ahler moduli space, some redefinitions are needed from
the $\cN=2$ counterparts. We will give the precise definitions of
the moduli in sections \ref{noscaleIIB} and \ref{noscaleIIA}.

The superpotential for SU(3) compactifications has been computed
in \cite{GLW,iman} and reads simply, \beq \label{Wgen} W= \int
\langle \Phi_1, d_H \, \Pi \rangle \ . \eeq

From (\ref{SU3}) and (\ref{Pi}) we observe that,
\begin{align}
&\textrm{O5/O9:} &  \quad \Phi_1&=\left(\frac{\mathcal{N}_J}{\mathcal{N}_{\Omega}}\right)^{1/2}\Omega\ ,  & \Phi_2&=i e^{-iJ} \ , \quad  &d_H\Pi&=F_3+ i e^{-\phi}d_HJ \ ,\label{go9}\\
&\textrm{O6:} & \quad \Phi_1&=e^{-iJ}\ , & \Phi_2&=\left(\frac{\mathcal{N}_J}{\mathcal{N}_{\Omega}}\right)^{1/2}\Omega \ ,  \quad  &d_H\Pi&=F+iC\RE
d_H\Omega \label{go6} \ ,
\end{align}
where
$C=e^{-\phi}(\mathcal{N}^{-1}_{\Omega}\mathcal{N}_J)^{1/2}=e^{-\phi_{(4)}}
\mathcal{N}_{\Omega}^{-1/2}$.
Substituting in (\ref{Kgen}) and (\ref{Wgen}) then leads to the familiar
expressions for type IIA and IIB,
\begin{align}
&\textrm{O5/O9:} & & K=-\textrm{log}\left[-i\int \Omega\wedge \overline \Omega\right]-\textrm{log }e^{-4\phi_{(4)}}\label{KgenIIB}\\
& & & W=\int \Omega\wedge(F_3+ie^{-\phi}dJ) \ , \label{super9o}\\
&\textrm{O6:} & & K=-\textrm{log}\left[\frac{4}{3}\int J\wedge J\wedge J\right]-\textrm{log }e^{-4\phi_{(4)}}\label{KgenIIA}\\
& & & W=\int [e^{iJ}\wedge F]_6+iC\int \RE
\Omega\wedge(H+idJ)\ ,
\end{align}
where we have performed a partial integration in order to derive the
IIA superpotential and a K\"ahler transformation to eliminate
an extra factor $(\mathcal{N}_{\Omega}/\mathcal{N}_J)^{1/2}$ from
the IIB superpotential. In what follows we will use the notation $G\equiv d_H\Pi$.

A generic K\"ahler potential $K$ and superpotential $W$ define a
potential $V$ equal to \beq \label{Vgen} V=e^K\left( \sum_{i,j}
K^{i\bar{\jmath}}D_iWD_{\bar{\jmath}}\overline{W}-3|W|^2\right)
\eeq where $i,j$ run over all the $\N=1$ moduli, and $D_i W =
(\partial_i + K_i) W$ (with $K_i=\partial_i K$). If (i) the
K\"ahler potential for a subset of moduli $\{ \tilde{\imath} \}$
satisfies a no-scale condition
\cite{noscale}, \beq \label{Knoscale} \sum_{\tilde{\imath}}
K^{\tilde{\imath} \bar{\tilde{\jmath}}} \partial_{\tilde{\imath}}
K \partial_{\bar{\tilde{\jmath}}} K = 3 \ , \eeq (ii) there are no mixed
terms $K^{\tilde{\imath} \bar{\jmath}}$ in the inverse of the
K\"ahler metric, and (iii) the superpotential is independent of
the moduli $\tilde{\imath}$, then the negative piece $-3 |W|^2$ in
the potential is cancelled, and the resulting potential is
positive definite, \beq \label{Vnoscale} V=e^K \sum_{i, j \neq
\tilde{\imath}} K^{i\bar{\jmath}}D_iWD_{\bar{\jmath}}\overline{W}
\ . \eeq This potential has an absolute minimum at $V=0$ when $D_i
W=0$, for all $i\neq \tilde{\imath}$. At this no-scale minimum
supersymmetry is broken by the F-terms of the moduli $\tilde{\imath}$,
since $D_{\tilde{\imath}} W= K_{\tilde{\imath}} W \neq 0$. We will
see in sections \ref{noscaleIIB} and \ref{noscaleIIA} that there
are several choices for the subset $\{ \tilde{\imath}\}$ for the
K\"ahler potential (\ref{Kgen}).

\subsection{D-branes on SU(3) structure manifolds} \label{genbranes}
We consider D-branes extended in 4d Minkowski space-time, wrapping
an internal cycle $\Sigma$. The world-volume combination
$\cF=F-P_\Sigma [B]$ (with $P_\Sigma$ the projection along
$\Sigma$) should satisfy the Bianchi identity $d{\cal
F}=-P_{\Sigma}[H]$.  We denote a brane by the pair $(\Sigma, {\cal
F})$.

Supersymmetric ``generalized cycles'' $(\Sigma, {\cal F})$ have to
satisfy the D-flatness and F-flatness conditions. The former reads
\cite{mar1}, \beq \label{Dflat} {\cal D}(\Sigma, {\cal
F})=P_\Sigma [e^{2A-\phi} \IM \Phi_2] \wedge e^{\cal
F}|_{\rm{top}}=0 \eeq where $\Phi_2= \Phi_{-} (\Phi_+)$ in IIA
(IIB), is the non-integrable pure spinor in an $\cN =1$ vacuum.
Notice however that (\ref{N=1vacua}) implies that $e^{2A-\phi} \IM
\Phi_2$ is closed on the supersymmetric vacuum.

The F-flatness conditions,
 which
can be derived from the superpotential (\ref{supgen}) below, read
\beq \label{Fflat}
F_m (\Sigma, {\cal F})= P_\Sigma [ e^{3A -\phi} (\iota_m + g_{mn} dy^n \wedge ) \Phi_1] \wedge
 e^{\cal F}|_{\rm{top}}=0
\eeq
with $\Phi_1=\Phi_{+} (\Phi_-)$ for IIA (IIB), the integrable pure spinor, and $\iota_m$ denotes a contraction along $\del_m$.

The F-flatness conditions imply that D-branes wrap generalized
complex submanifolds  $(\Sigma, {\cal F})$
\cite{gualtieri,Koerber,mar0}, which means that their generalized
tangent bundle \beq T_{(\Sigma,{\cal F})}=\{ v+ \xi \in T\oplus
T^* |_\Sigma : \iota_v {\cal F}=P_\Sigma [\xi]\} \eeq is stable
under the integrable generalized complex structure associated to
$\Phi_1$  (i.e. if we denote ${\cal J}_1$ this generalized complex
structure, ${\cal J}_1 X \in T_{(\Sigma,{\cal F})}$, $\forall X
\in  T_{(\Sigma,{\cal F})}$). In type IIB, $\Phi_1=\Phi_-$ is
proportional to $\Omega$, which defines a complex structure, and
therefore the generalized complex submanifolds are complex
submanifolds and ${\cal F}$ is (1,1). In type IIA, $\Phi_1
=\Phi_+$ is proportional to $e^{-iJ}$, which defines a symplectic
structure, and therefore the complex submanifolds wrapped by
D6-branes are special Lagrangian, and ${\cal F}=0$. We will see
this in more detail in sections \ref{muIIB} and \ref{muIIA}. The
D-term conditions are stability conditions for the D-brane.

The deformations of the cycle are sections of the generalized
normal bundle $N_{(\Sigma,{\cal F})}=(T_M \oplus T^*_M)|_{\Sigma}
/ T_{(\Sigma,{\cal F})}$ \cite{mar1}. Given a metric on the
manifold, we can split $T_M=T_\Sigma + T^\bot_\Sigma$. A section
of the generalized normal bundle is of the form
$X_{\perp}=(v_{\perp},a)$, where $v_\perp \in
\Gamma(T^\bot_\Sigma)$ generates the deformations of the cycle
$\Sigma$, while the deformations of the gauge field are $\delta
{\cal F}=da - P_{\Sigma}[\iota_{v_\perp} H]$. The last term
insures than under deformations of $\Sigma$, the Bianchi identity
$d{\cal F}= - P_{\Sigma}[ H]$ still holds. Since the tangent
bundle is stable under the integrable generalized complex
structure ${\cal J}_1$, the latter induces a natural a complex
structure on $N_{(\Sigma,{\cal F})}$. This implies that the
holomorphic generalized normal vectors, which are associated to
the four-dimensional chiral fields on the brane, are $Z=(1-i \cJ)
X_{\perp}$.

For A-branes,  ${\cal J}_1$ corresponds to the B-transformed of the
symplectic structure $J$, given in (\ref{gensympl}). The 1-form
part of the holomorphic generalized normal vectors is therefore
given by $(1+ i B J^{-1}) \left(a+(B+iJ) v_{\perp} \right)$. (The
vector part is just $-i J^{-1} (1+ i B J^{-1})^{-1}$ times the
1-form part). Furthermore, for supersymmetric configurations the
$H$ field is zero, and therefore $a$ represents pure gauge
transformations of the world-volume field-strength. The
holomorphic fields on the brane, $\phi_i$, are consequently given
by \beq \label{Afields} \phi_i = [A + (B+ iJ) \, v_{\perp}]_i \ ,
\quad \textrm{type IIA} \eeq

For B-branes ${\cal J}_1={\cal J}_-$, given in (\ref{gencompl}),
and corresponds to an ordinary complex structure $I$. $H$ is also
zero for supersymmetric O5/O9 configurations. Here it is very easy
to see that the holomorphic generalized tangent vectors are given
by the holomorphic normal vectors and the holomorphic 1-form gauge
field, namely \beq \label{Bfields} \phi_i=[(1+i \,  I^T) A]_i \ ,
\qquad \phi^i=[(1-i \, I) \, v_{\perp}]^i \  , \quad \textrm{type
IIB} \eeq

The geometrically induced $\mu$-terms can be computed from the
D-brane superpotential. For a D-brane wrapping the cycle $(\Sigma,
{\cal F})$, the superpotential is \cite{mar1}, \beq \label{supgen}
W=\int_{\cal B} P_{\cal B}[e^{3A-\phi} \Phi_1] \wedge e^{\cal
\tilde F} \eeq where $({\cal B}, {\cal \tilde F})$ is a chain
whose boundaries are a fixed generalized cycle $(\Sigma_f,{\cal F}_f)$
and $(\Sigma,{\cal F})$. As we will see in sections \ref{muIIA},
\ref{muIIB}, the D-brane superpotential is holomorphic in the
D-brane fields (\ref{Afields}) and (\ref{Bfields}).

\section{Type IIB compactifications with O9/O5-planes} \label{bulkIIB}

In this section we use the K\"ahler potential and bulk and brane
superpotentials reviewed in the previous section to find no-scale
supersymmetry breaking vacua for type IIB compactifications with
O9 and O5-planes, as well as the geometrically induced $\mu$-terms
on D9 and D5-branes. Backgrounds preserving this sort of
supersymmetries have been described for example in~\cite{GMPT1} under the
label of type C solutions. Perhaps, the best known representative is
the background constructed by Chamseddine and Volkov \cite{CV},
whose AdS/CFT interpretation was given by
Maldacena-Nu\~{n}ez~\cite{hepth0008001}. Here we will consider the
possible $\mathcal{N}=0^*$ no-scale deformations of this kind of
backgrounds.

\subsection{No-scale vacua} \label{noscaleIIB}

Let us write again the bulk superpotential for this type of
compactification, given in (\ref{super9o}),
\begin{equation}
W= \int \Omega \wedge (F_3+ie^{-\phi}dJ) \label{super9}
\end{equation}
In order to extract the maximum information from it, it is
convenient to decompose it into irreducible representations of the
underlying SU(3)-structure. The complex 3-form
$G= F_3+ie^{-\phi}dJ$,
transforming in a $\mathbf{20}=\mathbf{10}\oplus\overline{\mathbf{10}}$
of $SU(3)$, decomposes as
\begin{equation}
G=G^+ + G^- \quad , \quad *_6G^{\pm}=\pm i G^{\pm}\ ,
\end{equation}
and
\begin{align}
G^{+}&=\frac{3}{2}\frac{\mathcal{N}_J}{\mathcal{N}_{\Omega}}G_{(1)}^{+}\overline{\Omega}+G_{(3)}^{+}\wedge J + G_{(6)}^{+}\ , \nonumber \\
G^{-}&=\frac{3}{2}\frac{\mathcal{N}_J}{\mathcal{N}_{\Omega}}G_{(1)}^{-}\Omega+G_{(3)}^{-}\wedge J + G_{(6)}^{-}\ , \label{su3g3}
\end{align}
with $G^{\pm}_{(1)}$ a complex zero form in the $\mathbf{1}$ of
$SU(3)$, $G^{+}_{(3)}$ ($G^{-}_{(3)}$) a complex (0,1)-form
((1,0)-form) in the $\bar{\mathbf{3}}$ ($\mathbf{3}$), and
$G^{+}_{(6)}$ ($G^{-}_{(6)}$) a complex primitive (2,1)-form
((1,2)-form) in the ${\mathbf{6}}$ ($\bar{\mathbf{6}}$). We
summarize in Appendix \ref{su3decomp} the different
representations and forms arising in the decomposition of $G$.

As already pointed out in section \ref{su3def}, on an SU(3)-structure
manifold there are no globally defined 1-forms. $G_{(3)}^{\pm}$
and $\mathcal{W}_5$, laying in topologically trivial
representations, encode information relative to the backreaction of
the fluxes and branes, and in the probe limit ($A\to 0$)
$F_3\wedge J=\mathcal{W}_4=\mathcal{W}_5=0$. For the moment we
concentrate on this limit, and latter on we will extend the
solution to the full one with finite warping.

Plugging (\ref{su3g3}) into (\ref{super9}), we get
\begin{equation}
W=\frac{3}{2}\frac{\mathcal{N}_J}{\mathcal{N}_{\Omega}}\int
G^+_{(1)} \Omega\wedge \overline{\Omega}=
12i\mathcal{N}_JG^+_{(1)}\ . \label{supervac}
\end{equation}
For purely imaginary self-dual (ISD) fluxes, which we will see is
a required condition, $G^+_{(1)}=e^{-\phi}\mathcal{W}_1$. A
non-vanishing gravitino mass therefore generically requires the
torsion class $\mathcal{W}_1$ to be non-vanishing. For vacua with
spontaneously broken supersymmetry, the generalized almost complex
structure is thus not integrable.

For simplicity, in what follows we assume $b_-^{(1,1)}=0$\footnote{$b^{(1,1)}_-$
is the number of odd (1,1) forms in the expansion in ``light modes'' (for more details, see \cite{GLW,iman,KPM})  In the
Calabi-Yau case, this would be the number of odd harmonic (1,1)-forms. For $b_-^{(1,1)} \neq 0$,
the expansion in moduli gets slightly more complicated (see \cite{hepth0403067}). We do not give it since the no-scale condition needs
$b_-^{(1,1)}=0$.}. The moduli space consists in the
complex structure moduli, $U^k$, $k=1,...b^{(2,1)}_+$, the complexified K\"ahler deformations, $T^a$,
defined from the expansion of the form in (\ref{Pi}), which in this case is
\beq \label{Ta}
\Pi=C_2 + \, ie^{-\phi} J= iT^a
\omega_a \ , \qquad a=1,...,b^{(1,1)}_+ \ ,
\eeq
and the axio-dilaton moduli, $S=e^{-\phi}\mathcal{N}_J+iC_{\mu
\nu}$. In terms of these, the K\"ahler potential (\ref{KgenIIB})
splits as
\begin{equation}
K_{\Omega}=-\textrm{log}[8\mathcal{N}_{\Omega}] \quad , \quad K_{J}=-\textrm{log}[8e^{-3\phi}\mathcal{N}_{J}] \quad , \quad K_S=-\textrm{log}(S+S^*)\ ,
\end{equation}
where ${\cal N}_\Omega$ and $e^{-3\phi} {\cal N}_J$ should be understood as functions of $U^k$ and
$T^a$ respectively.

Let us compute the F-terms coming from (\ref{super9}). These are
proportional to the covariant derivatives, $D_kW\equiv
\partial_kW+W\partial_kK$, with respect to the above moduli,
\begin{align}
D_{U^k}W&=-\int \chi_k\wedge G^{-}_{(6)}\ , \\
D_{T^a}W&=12ie^{\phi}G_{(1),a}^+\mathcal{N}_J\ , \\
D_SW&=K_SW\ ,
\end{align}
where the set of primitive $(2,1)$ forms $\chi_k$ is defined through,
\begin{equation}
\chi_k\equiv \frac{\partial_{U^k}\mathcal{N}_{\Omega}}{\mathcal{N}_{\Omega}}\Omega-\partial_{U^k}\Omega \ ,
\end{equation}
and we have expanded $G_{(1)}\mathcal{N}_J$ as
\begin{equation}
G_{(1)}\mathcal{N}_J=(G_{(1),a}\mathcal{N}_a+G_{(1)}\mathcal{N}_{J,a})T^{a}+\overline{F}_{(1)}\mathcal{N}_J
\ , \label{gexpan}
\end{equation}
with $\overline{F}_{(1)}$ the scalar component of the $F_3$
background, as defined in (\ref{f3desc}).
For a purely ISD background,
$G_{(1),a}^+=e^{-\phi}\partial_{T^a}\mathcal{W}_1$.
In what
follows we define $\mathcal{W}_{1,a}\equiv
\partial_{T^a}\mathcal{W}_1$, $\mathcal{N}_{J,a}\equiv
\partial_{T^a}\mathcal{N}_{J}$ and $\mathcal{N}_{\Omega,k}\equiv
\partial_{U^k}\mathcal{N}_{\Omega}$ to simplify the notation.

For $\mathcal{N}=1$ supersymmetric vacua the F-terms have to vanish,
and this requires
\begin{align}
&\mathcal{W}_1=F_{(1)}=0\ , \label{susyw1} \\
&\mathcal{W}_3=-e^{\phi}*_6F_{(6)}\ , \label{susyw3}
\end{align}
in agreement with the conditions coming from (\ref{N=1vacua}).

We may think about relaxing these conditions in order to obtain
no-scale solutions with spontaneously broken supersymmetry. For
that aim, as it will be clear below, it is convenient to take the
internal manifold to be the fibration of a complex 2-cycle
$\Sigma_2$ over a four dimensional base $B$. The K\"ahler form
splits accordingly,
\begin{equation}
J=J_B+J_{\Sigma_2} \ , \label{jdecomp}
\end{equation}
with $J_B\wedge J_B\wedge J_{\Sigma_2}=2\mathcal{N}_J$. In
general, the 2-cycle may not be trivially fibered over the base
$B$, which in terms of torsion classes means $dJ_{\Sigma_2}\neq
0$. In addition, whenever compatible with the $\Sigma_2$
fibration, the base manifold itself may have non-trivial intrinsic
torsion, $dJ_B\neq 0$, as long as  $dJ_B \wedge \Omega=0$.
This ensures that the superpotential does not
depend on the K\"ahler moduli of the base, $T^{\tilde a}$.

Taking a non-vanishing $G_{(1)}^+$, but independent of the
K\"ahler moduli of the fiber, $T^{b}$, leads to
\begin{align}
D_SW&=K_SW\ , \\
D_{U^k}W&=-\int \chi_k\wedge G^{-}_{(6)}\ , \label{susynosca}\\
D_{T^a}W&=\begin{cases}K_{T^{\tilde a}}W  & \textrm{for } T^{\tilde a} \\
0 & \textrm{for }  T^{b} \end{cases}\ .
\end{align}
Therefore, imposing $D_{U^k}W=0$, the negative piece of the
scalar potential is exactly cancelled by the non-vanishing
F-terms, and we get a positive definite no-scale potential of the
form (\ref{Vnoscale}), where the sum runs over $i= T^{b},
U^k$.

In terms of the torsion classes and the 3-form RR background, the
above conditions for a no-scale supersymmetry breaking vacuum read,
\begin{align}
&\mathcal{W}_1=e^{\phi}F_{(1)}\ , \label{1nosca}\\
&\mathcal{W}_3=-e^{\phi}*_6F_{(6)}\ ,\label{6nosca}
\end{align}
where  to get the first equation we have used (\ref{gexpan}) and the fact that
$\del_{\tilde a} W=0$ implies
$\mathcal{N}_{J,\tilde a}G^+_{(1)}=-\mathcal{N}_{J}G^+_{(1),\tilde a}$. $G$ is therefore a
purely ISD form.

Furthermore, $\mathcal{W}_2$  will generically be different from
zero. Indeed, for $G$ purely ISD, we can reexpress the
superpotential as,
\begin{equation}
W=2ie^{-\phi}\int \Omega\wedge dJ=-2ie^{-\phi}\int d\Omega\wedge J
\ .
\end{equation}
Decomposing $J$ as in (\ref{jdecomp}), taking
$T^{\tilde a}\partial_{\tilde a} W=0$ and using $\mathcal{W}_2\wedge J\wedge J = 0$,
we obtain
\begin{equation}
 2\mathcal{W}_1 J_B\wedge J_B\wedge J_{\Sigma_2} - \mathcal{W}_2\wedge
J_B\wedge J_{\Sigma_2} = 0  \ .
\end{equation}
This completely determines $\mathcal{W}_2$ in terms of $\mathcal{W}_1$,
resulting in,
\begin{equation}
\mathcal{W}_2=2\mathcal{W}_1(J_B-2J_{\Sigma_2})\ . \label{w2}
\end{equation}

Some comments are in order. First, notice that the supersymmetry
breaking is mediated through F-terms associated to the moduli in
the expansion of $\Pi$ ($S$ and $T^{\tilde a}$). Those descend from $\mathcal{N}=2$
hypermultiplets spanning a quaternionic manifold. The same
situation occurs in conventional Calabi-Yau compactifications of type IIB
with O3-planes and 3-form fluxes. More
concretely, when $\Sigma_2$ is a 2-torus and $dJ_B=0$, both kind
of setups can be related by two T-dualities on
$\Sigma_2$.\footnote{For a supersymmetric version of this setup
see \cite{hepth0406001}.} By a slight abuse of language, we will
denote this type of breaking, characterized by an ISD 3-form
(or more generically by an ISD polyform) with a non-vanishing
SU(3) singlet component, as ``no-scale quaternionic breaking''. Further
examples of this type of breaking will appear in section
\ref{IIAquat} for type IIA orientifolds.

Finally, let us comment on the warp factor. We have argued that the
superpotential for SU(3)-structure compactifications does not
contain the effects of the warping, as these are encoded in
topologically trivial representations. Rather, they appear as
corrections to the K\"ahler potential of the 4d effective
theory~\cite{warp}. Since the non-supersymmetric piece of the
background is exclusively contained in the SU(3) invariant term,
as can be read from (\ref{1nosca}), we do not expect this
deformation to mix with quantities transforming in vector
representations. The latter should therefore satisfy the same
relations than in the supersymmetric case, given by
(\ref{N=1vacua}),
\begin{equation}
2i\mathcal{W}_5^*=-e^{\phi}F_{(3)}=-2i\bar{\partial}A=-i\overline{\partial}\phi\
. \label{warprel}
\end{equation}

Further support to this idea comes from the analysis of the RR
tadpoles. The experience with ordinary flux compactifications and
open/closed string duality tells us that the backreacted geometry
can be alternatively characterized by the induced charges in the
bulk. The relevant piece of the ten dimensional action
is~\cite{hepth0602089},
\begin{multline}
\int C_6\wedge dF_3=-i\int C_6\wedge dG = \int C_6\wedge
\left[\frac{3}{2}\frac{\mathcal{N}_Je^{-\phi}}{\mathcal{N}_{\Omega}}\left(|\mathcal{W}_1|^2J\wedge
J+\mathcal{W}_1\overline{\mathcal{W}}_2\wedge J\right)+e^{-\phi}d*_6\mathcal{W}_3\right]\\
=
\frac{9}{2}\frac{\mathcal{N}_Je^{-\phi}|\mathcal{W}_1|^2}{\mathcal{N}_{\Omega}}\int
C_6\wedge J_B\wedge J_B\ + \ e^{-\phi}\int C_6\wedge d*_6\mathcal{W}_3 \ ,
\end{multline}
where we have made use of (\ref{w2}) for the last equality. The
intrinsic torsion therefore induces a non-vanishing charge of
D5-brane along $\Sigma_2$, and the backreacted geometry is
expected to lay within the same class than the one produced by a
stack of D5-branes wrapping $\Sigma_2$. The latter indeed can be
shown to satisfy equation
(\ref{warprel})~\cite{Horowitz:1991cd,hepth9412184}.

For the sake of clarity, let us now discuss a particular example
of no-scale quaternionic breaking with O9/O5-planes.

\subsection{Example: $K3\times T^2$ fibration} \label{sec:exIIB}

Consider a compact nilmanifold with tangent 1-forms $e^i$
satisfying the equations,
\begin{align}
&de^1=de^2=de^4=de^5=0\quad , \quad de^6=e^4\wedge e^5 \quad , \nn \\
&de^3=e^4\wedge e^5-e^1\wedge e^5+e^2\wedge e^4\ . \label{tdual2}
\end{align}
This corresponds to a $T^2$ fibration over a factorizable $T^4$
spanned by the coordinates $x^1$, $x^2$, $x^4$ and $x^5$. Notice
that this set of equations is invariant under a $\mathbb{Z}_2$
discrete symmetry reversing the coordinates of the base. We take
the orientifold generated by the combined action
$\Omega_P\mathbb{Z}_2$, leading to a set of O5-planes with total
charge of 16 (in D5-brane units) wrapping the $T^2$ fiber. This
construction can be understood as the orbifold limit of a
$K3\times T^2$ fibration (see \cite{cvetic} for related
constructions).

The internal metric in (\ref{metric}) is given by
\begin{equation}
ds_6=\frac{e^{-2A}}{u}\left( \tilde t_1|e^1+iu e^4|^2+ \tilde t_2|e^2+iu e^5|^2\right)+\frac{e^{2A}}{u} t|e^3+iue^6|^2\
, \label{metrick3t2}
\end{equation}
with $\tilde t_i$, $t$ the real parts of, respectively,
the K\"ahler moduli of the base and the fiber\footnote{We will always denote the real
part of a field with the same letter in lowercase.}
and $u$ is the overall complex structure modulus, that is fixed to a real value in the solution
(i.e., the complex structure axions are zero). Here,
$e^i=dx^i$ for $i=1,2,4,5$ and
\begin{equation}
e^3=dx^3+x^4dx^5-x^1dx^5-x^4dx^2 \ , \quad e^6=dx^6+x^4dx^5 \ .
\end{equation}

$J$ and $\Omega$ are given by
\begin{align}
J&=J_B+J_{T^2}=-e^{-2A}(\tilde t_1 e^1\wedge e^4+\tilde t_2 e^2\wedge
e^5)-e^{2A} t\ e^3\wedge e^6\ , \label{jex1}\\
\Omega&=e^{-A}(e^1+iu e^4)\wedge(e^2+iu e^5)\wedge(e^3+iue^6)\
.\label{omgex1}
\end{align}
Notice in particular that $dJ_B=0$, as corresponds to a CY${}_2$
manifold, and the model can be related to an ordinary  $T^6$
orientifold with O3-planes and ISD 3-form flux by T-dualizing
along $x^3$ and $x^6$.

From (\ref{jex1}) and (\ref{omgex1}) we extract the torsion
classes,
\begin{align}
\mathcal{W}_1&=-\frac{e^{3A}}{6\tilde t_1\tilde t_2}(3iu+1) \ , \\
\mathcal{W}_2&=-\frac{e^{3A}}{3\tilde t_1\tilde t_2}(3iu+1)(J_B-2J_{T^2})\ , \\
\mathcal{W}_3&=\frac{e^{2A}}{8}\frac{t}{u^3}(u+i)(z^1\wedge
z^2\wedge \bar z^3+ z^1\wedge \bar z^2\wedge z^3+\bar z^1\wedge
z^2\wedge z^3) + c.c \ \\
& \quad - 2 dA\wedge (J_B-J_{T^2}) \ , \\
\mathcal{W}_4&=0 \ , \\
\mathcal{W}_5&=-\partial A \ ,
\end{align}
with $z^a\equiv e^a+iue^{a+3}$ the holomorphic 1-forms, and we are
defining ${\cal N}_J$ and $\NO$ by (\ref{NJNO}) in the limit $A\to 0$.
Observe that $\mathcal{W}_1$ does
not depend on $t$, so the manifold has a suitable structure
to support a no-scale solution of the type described in the previous
section. For that, eqs. (\ref{1nosca}), (\ref{6nosca}) and
(\ref{warprel}) dictate the exact expression of the 3-form
background,
\begin{equation} \label{F3ex}
g_s F_3=-\frac{
t}{8 u^3}[(1-3iu)\Omega+(1-iu)(z^1\wedge z^2\wedge \bar
z^3+z^1\wedge \bar z^2\wedge z^3+ \bar z^1\wedge z^2\wedge
z^3)]+e^{2A}*_4d(e^{-4A})+c.c
\end{equation}
with $*_4$ the hodge star in the base, and $g_s$ the VEV of $e^{\phi}$, i.e.
we use $e^\phi=g_s e^{2A}$. Supersymmetry is broken by the F-terms of $S$, $T_1$ and
$T_2$, proportional to $\mathcal{W}_1$, and the cosmological
constant vanishes at tree level, accordingly to the no-scale
structure. Finally, the Bianchi identity for $F_3$ determines the charge
of $D5_{T^2}$-brane induced by the flux,
\begin{equation}
dF_3=\frac{e^{4A}}{2 g_s \tilde t_1\tilde t_2}\left(\frac{t}{u^3}(1+3u^2)+e^{-2A}\nabla_{B}^2(e^{-4A})\right)J_B\wedge J_B\ .
\end{equation}

\subsection{Geometrically induced $\mu$-terms on twisted tori in IIB} \label{muIIB}

In the above compactifications, besides the flux, there are generically
also D5 and D9-branes wrapping respectively complex 2-cycles and the whole space.
These are required to cancel the global negative RR charge induced by the
orientifold planes, whenever it is not cancelled completely by the flux.
The deformations of these branes are
parameterized by the holomorphic normal vectors, $\phi^i$, and the
holomorphic 1-form gauge fields, $\phi_i$, given in equation
(\ref{Bfields}). In a realistic compactification these would be
identified with the supersymmetric partners of the matter fields,
i.e. with the squarks and sleptons. The pattern of soft
supersymmetry breaking terms can be thus determined by the F-terms
together with the possible $\mu$-terms for $\phi^i$ and $\phi_i$.

The superpotential (\ref{supgen}) constitutes a simple way for
computing the $\mu$-terms in a given class of compactification.
For the particular case of D9 and D5-branes, it reduces
to~\cite{mar1},
\begin{align}
W_{D9}&=\int \Omega\wedge \omega_3\ ,\label{mud9} \\
W_{D5}&=\sum_i\int_{\mathcal{B}_i}\Omega\ , \label{mud5}
\end{align}
where $\omega_3$ is the Chern-Simons 3-form,
\begin{equation}
\omega_3=\textrm{Tr}\left(A\wedge dA+\frac{2}{3}A\wedge A\wedge
A\right)\ ,
\end{equation}
and $\{\mathcal{B}_i\}$ the set of 3-chains generated by all
possible infinitesimal deformations of the generalized complex
2-cycle which the D5-brane wraps. Alternatively, (\ref{mud9}) can
be anticipated by arguments of anomaly cancellation, since
coupling 10d super Yang-Mills to the bulk supergravity
\cite{Chamseddine:1980cp,Bergshoeff:1981um,Chapline:1982ww}
requires $F_3\to F_3+\omega_3$ in (\ref{super9}), giving rise to
(\ref{mud9}).

Performing the integral (\ref{mud5}) requires a precise knowledge
of the embedding of the
D5-brane in the geometry of the internal manifold. Moreover, the
zero modes of $\phi^i$ and $\phi_i$ may have non-constant profiles
on the compact directions. For all this, we restrict here to the
particular case of twisted tori.

A twisted torus is an homogeneous parallelizable manifold with a set of globally defined
1-forms $e^a$. These are not closed, but satisfy the Maurer-Cartan equations
\begin{equation}
de^a=\frac{1}{2}f^a_{bc}e^b\wedge e^c \ , \label{torsion}
\end{equation}
for constant $f^a_{bc}$. Imposing $d^2 e^a=0$, requires the constants to satisfy
Jacobi identities
\begin{equation}
f^a_{[bc}f^g_{d]a}=0 \ . \label{metcons}
\end{equation}
$f^a_{bc}$ are therefore structure constants of a Lie algebra of a
group $G$. The twisted torus is the manifold $G/\Gamma$, where
$\Gamma$ is a set of discrete identifications. For $f^a_{bc}=0$,
these are of the form $x^a\cong x^a+k^a$ for some constants $k^a$,
while in the case of nonzero structure constants some of these
identifications are ``twisted'' (for example, if $f^3_{12}=h$ and
the rest are zero, then one can identify $x^1 \cong x^1 + k^1$,
$x^2 \cong x^2+k^2$, $x^3 \cong x^3 - k^1 h x^2$).

Since twisted tori are manifolds of trivial structure, as they are parallelizable,
one can globally define many SU(3) structures on them.
They are defined by the following pairs of $\Omega$ and $J$
\begin{equation}
\Omega =z^1\wedge z^2\wedge z^3 \quad , \quad J = j_{m\bar n}z^m\wedge \bar z^n \ , \label{omtwist}
\end{equation}
where $j_{m\bar{n}}=-(j_{n\bar{m}})^*$ and in the basis of holomorphic
1-forms, $z^m\equiv e^{m}+iU^m{}_ne^{n+3}$, $m,n=1,2,3$, the metric reads
$g_{m\bar{n}}=-i j_{m\bar{n}}$. For completeness, we give the torsion classes in terms of the
structure constants in Appendix \ref{torsiontori}.

A stack of D9-branes wrapping the entire volume of the twisted
torus will contain three complex Cartan moduli
$\phi_m=A_m+iU_m{}^nA_{n+3}$. From (\ref{mud9}) one may extract
then the $\mu$-terms for the light modes,
\begin{equation}
W_{\mu,D9}=\frac{i\mathcal{N}_\Omega}{2}g_{m\bar o}g_{p\bar q}\epsilon^{\bar s \bar r \bar o}f^{\bar{q}}_{\bar{s}\bar{r}}\phi^m\phi^p
\ , \label{finmud9}
\end{equation}
where $\phi^m=g^{m \bar n}\phi_{\bar n}$ and $\epsilon^{\bar 1 \bar 2 \bar 3}= \epsilon_{123}=-i$.

Similarly, for a stack of D5-branes wrapping the complex 2-cycle
$\Pi=a_{i \bj}[z^i\wedge \bar z^j]$, there are two normal moduli
$\phi^m$, $m=1,2$, plus a single Cartan moduli $\phi$. Performing
a change of basis, $\{z^1,z^2,z^3\}\to \{\tilde z^1,\tilde z^2,\tilde z^3\}$,
such that in the new basis $\Pi=[\tilde z^3\wedge \bar{\tilde z}^3]$, these moduli
can be identified respectively with the normal coordinates $\tilde z^1$ and $\tilde z^2$
and with the gauge bundle along $\tilde z^3$.

From (\ref{mud5}) then we extract,
\begin{equation}
W_{\mu,D5}=\frac{i}{2}\epsilon_{3jk} f^k_{\bar{3} m}\phi^m\phi^j\ ,\label{finmud5}
\end{equation}
where the indices now refer to the new complex coordinates $\tilde{z}^i$.

The complexified structure constants defining the topology of the
twisted torus can be therefore arranged according to the
holomorphicity/antiholomorphicity of their indices. Thus,
$f^a_{b\bar{c}}$ gives rise to $\mu$-terms for geometric moduli of
D5-branes, whereas $f^a_{bc}$ corresponds to $\mu$-terms for the
Cartan moduli of D9-branes. On top of this, $f^a_{\bar b\bar c}$
controls the amount of supersymmetry breaking, as derived from
(\ref{w1twist}).

Notice that the superpotentials (\ref{finmud9}) and
(\ref{finmud5}) involve structure constants with complex indices.
In writing them in terms of the ones with real indices,
non-holomorphic pieces in the complex structure moduli appear. Strictly
speaking, only the holomorphic terms correspond to infinitesimal
supersymmetric deformations of the calibrated
branes~\cite{mar1,mar2}, whereas the non-holomorphic pieces can be
generically traced back to $\phi^i\phi^j$ couplings in the
K\"ahler potential~\cite{0501139}, giving effective contributions
through the Giudice-Masiero mechanism~\cite{giudice} in vacua with
supersymmetry spontaneously broken. We will come back to this
issue in section \ref{soft}. Besides, the superpotential
(\ref{mud5}) contains also antiholomorphic terms in the brane
moduli, proportional to the structure constants $f^q_{\bar s \bar
r}$. These appear when the bulk almost complex structure is not
integrable, and correspond again to non supersymmetric
deformations of the branes. Therefore, these terms have to be
discarded.

\TABLE{\begin{tabular}{c||cccc}
&D9&D5${}_1$&D5${}_2$&D5${}_3$\\
\hline $u_1 \mu_{11}$& $\tilde f^{\bar{1}}_{\bar{2}\bar{3}}$ & $0$ & $\tilde f^3_{1\bar{2}}$ & $\tilde f^2_{\bar{3}1}$ \\
$u_2 \mu_{22}$ & $\tilde f^{\bar{2}}_{\bar{3}\bar{1}}$ & $\tilde f^3_{\bar{1}2}$ & $0$ & $\tilde f^1_{2\bar{3}}$ \\
$u_3 \mu_{33}$ & $\tilde f^{\bar{3}}_{\bar{1}\bar{2}}$ & $\tilde
f^2_{3\bar{1}}$ & $\tilde f^1_{\bar{2}3}$ & $0$
\end{tabular}
\caption{Supersymmetric torsion induced $\mu$-terms for D9 and
D5${}_i$ branes wrapping the $i$-th $T^2$ in a factorizable
twisted torus.}\label{mususy} }

In order to match the result for the D7-brane effective
$\mu$-term~\cite{hepth0408036,0501139} in vacua where a T-dual
description is available, the matter fields have to be rescaled
accordingly. We summarize in table \ref{mususy} the torsion
induced $\mu$-terms for the different types of D-branes present in
a compactification on a factorizable $T^6$, where the structure
constants have one leg on each 2-torus. For convenience, we have
introduced the ``rescaled'' structure constants, ${\tilde
f}^{I}_{JK}$, defined as
\begin{equation}
\label{renorma} {\tilde f}^{I}_{JK}\equiv \frac{2u_Ju_K}{t_I}
f^{I}_{JK}\ .
\end{equation}
In terms of these, the normalized $\mu$-terms for factorizable
twisted tori are
\begin{equation}
W_{\mu,D9}=\frac{i}{2}\epsilon_{srp}u_q^{-1}{\tilde f}^{\bar
q}_{\bar s\bar r}(\phi^p)^2 \ , \qquad
W_{\mu,D5_p}=\frac{i}{2}\epsilon_{pjk}u_j^{-1}{\tilde f}^k_{\bar p
m}(\phi^m)^2\ . \label{finmud9b}
\end{equation}

Notice that the spectrum is much richer than for type IIB
orientifolds with O3-planes and 3-form fluxes, where only the
geometric moduli of the D7-branes can be stabilized by the fluxes
\cite{gorlich,cascales,hepth0408036,gomis}. Concretely both the
Cartan moduli of the D9-branes and the geometric moduli of the
D5-branes can be lifted by the intrinsic torsion. A simple
intuitive example is provided by a $D9$-brane wrapping the product
$S^3\times T^3$, with holomorphic vectors $z^m=e^m+i\hat e^m$. The
left-invariant 1-forms of the 3-sphere satisfy
\begin{equation}
de^m=\frac{1}{2}\epsilon_{mno}e^n\wedge e^o\ , \label{ejemplito}
\end{equation}
whereas the ones in the 3-torus are closed, $d\hat{e}^i=0$. Since
$h^{(1,0)}[S^3\times T^3]=0$, we do not expect 4d massless zero
modes coming from the gauge bundle. In fact, in terms of
holomorphic vectors one has from (\ref{ejemplito}), $f^{\bar
1}_{\bar 2\bar 3}=f^{\bar 2}_{\bar 3\bar 1}=f^{\bar 3}_{\bar 1\bar
2}=1/4$, and therefore all the scalars transforming in the adjoint
are indeed lifted from the massless spectrum by the torsion
induced $\mu$-terms, leading to pure $\mathcal{N}=1$ super
Yang-Mills in 4d.

\section{Type IIA compactifications with O6-planes} \label{bulkIIA}

Type IIA compactifications with O6-planes have been one of the
preferred setups for D-brane model building during the last decade
(for a recent review see e.g. \cite{marchesano}). The easiness for
accommodating chiral fermions in bifundamental representations
without breaking $\mathcal{N}=1$ supersymmetry, makes it the
perfect framework for embedding realistic gauge theories in string
theory. Recently, the possibility of adding closed string fluxes
to type IIA orientifold compactifications has been also considered
\cite{derendinger,kachru,GL,VZ}, resulting in the (perturbative)
stabilization of all the closed string (untwisted) moduli of the
compactification \cite{gyravets,cfi,marios}. In this section we
construct no-scale supersymmetry breaking solutions of
type IIA compactified on orientifolds of SU(3)-structure manifolds. We will
see that the resulting possibilities turn out to be richer than
for type IIB orientifold compactifications.

\subsection{No scale vacua} \label{noscaleIIA}

The superpotential (\ref{Wgen}) specialized to IIA
compactifications with O6 planes is \beq \label{superIIA}W=\int
\langle e^{-iJ}, F+iC\RE d_H\Omega \rangle \equiv \int \langle e^{-iJ}, G \rangle\ , \eeq
where the
pairing $\langle , \rangle$ is defined in (\ref{Mukai}),  $F$ in
(\ref{F10F6})-(\ref{fstr}) and $C$ is the compensator field defined below
(\ref{go6}). Similarly to the type IIB case, we can
decompose the ``flux'' $G=F+iC\RE d_H\Omega$ into ISD and IASD
parts under the combined action $* \lambda $, \beq G=G^+ + G^- \ , \qquad
*\lambda[G^{\pm}] = \pm i G^\pm \ . \label{ghodge}\eeq $G^\pm$ can
be decomposed in representations of SU(3) in the following way
\cite{gmpt2} \bea \label{Gplus} G^+ &=& \frac{{\cal
N}_{\Omega}}{{\cal N}_J} \, G^+_{(1)} e^{iJ} +
 G^+_{mn} \gamma^m  e^{-iJ} \gamma^n  +G^+_m \gamma^m \, \bar \Omega_3 + \tilde G^+_m  \Omega_3 \gamma^m \ , \nn \\
G^- &=& \frac{{\cal N}_{\Omega}}{{\cal N}_J} \, G^-_{(1)} e^{-iJ} +
 G^-_{mn} \gamma^m  e^{iJ} \gamma^n  + G^-_m \gamma^m \Omega_3 + \tilde G^-_m
 \bar \Omega_3 \gamma^m \ ,
\eea where \beq \label{gammas} \gamma^m \Phi^\pm= (dx^m \wedge +
g^{mn} \iota_n) \Phi^\pm \ , \qquad \Phi^\pm \gamma^m=\pm
(dx^m\wedge - g^{mn} \iota_n) \Phi^\pm \eeq for $\Phi^+ (\Phi^-)$
any even (odd) form. The first terms in these expressions are
singlets of the SU(3) structure, the second are in the ${\bf 8} +
{\bf 1}$, while the last two are respectively in the ${\bf 3}$ and
${\bf \bar 3}$ representations. Each term can be obtained by an
appropriate integral. For example, \beq G_{(1)}^+= \frac{i}{8
{\cal N}_{\Omega}} \int \langle e^{- iJ}, G \rangle \ , \qquad
G^+_{mn}=\frac{i}{32 {\cal N}_J} J_{mp} J_{nq} \int \langle
\gamma^p e^{ iJ} \gamma^q, G \rangle  \ . \eeq We give in Appendix
\ref{su3decomp} the expressions for the other components. Using
this decomposition, the superpotential (\ref{superIIA}) is \beq
\label{supervacIIA} W=-8i \NO G_{(1)}^+ \eeq In type IIA
compactifications with O6-planes \cite{GL,iman}, the moduli are
the complexified K\"ahler deformations $T^a$ from the expansion
\beq \label{Ta2} B+iJ=i \, T^a \, \omega_a \ , \qquad a=1,\ldots
,b^{2}_- \eeq and the combination of axions and complex-structure
deformations encoded in the form $\Pi$ in (\ref{Pi}), namely
 \begin{align} \Pi= C_3 + i C \RE
\Omega=& (\xi^{K} + i C \RE Z^{K}) \alpha_{K} -
(\tilde \xi_\lambda + i C \RE {\cal F}_{\lambda}) \beta^\lambda
\equiv  i(N^{K} \alpha_{K} - U_{\lambda} \beta^\lambda)\ , \nn \\
& K=0,\ldots , h \ , \quad \lambda=h+1,\ldots ,b^3_+-1
\label{spinorexpan}\end{align} where
the integer $h$ is basis
dependent, $(\alpha_{K}, \beta^\lambda)
\equiv (\alpha_0, \alpha_k, \beta^{\lambda})$ are even 3-forms, paired symplectically with the odd 3-forms
$(\alpha_\lambda, \beta^{K}) \in \Delta^3_-$, and ${\cal
F}_{\lambda}$ the derivative of the prepotential with respect to
$Z^{\lambda}$.

The K\"ahler potential for the complex structure, K\"ahler and
axion-dilaton $S$ is given in (\ref{Kgen}), and reads in this case
 \beq K_J= - \textrm{log}[8 {\cal N}_J] \ ,
\qquad  K_{\Omega, S}=-2\textrm{log}[C^2  {\cal N}_{\Omega}]\ ,
\eeq
where $\NJ$ and $C^2 \NO$ should be written in terms of the moduli $T^a$,
$N^K$ and $U_\lambda$. The orientifold projection selects a privileged choice of the
symplectic basis in (\ref{spinorexpan}) for which $h=0$, i.e. \beq
\Pi = i(N^0\alpha_0-U_{\lambda} \beta^{\lambda}) \label{pipri} \
,\eeq and $N^0=S=C\RE Z^0-i\xi^0$. In the large complex structure
limit\footnote{This setup is mirror to large volume
compactifications of IIB with O3/O7 planes in the case $b^2_-=0$.
We thank T. Grimm for pointing this out to us.}, ${\cal
F}=\frac{1}{Z^0} k_{abc} Z^a Z^b Z^c$, and the K\"ahler potential
for $S$ and $U_\lambda$, $\lambda=1\ldots b^3_+-1$, splits into
\beq K_{\Omega, S} = -\textrm{log} (S+ S^*) - 2 \textrm{log}
({\cal K}_{U_{\lambda}})\ , \label{kahlerlarge}\eeq where ${\cal
K}_{U_{\lambda}}= s^{3/2} \kappa_{\alpha \beta \rho} \tau^{\alpha}
\tau^\beta \tau^\rho $ and $\tau^{\lambda}\equiv \frac{C \RE
Z^{\lambda}}{C \RE Z^0}$ should be solved as a function of
$U_{\lambda}$. The last piece is of the no-scale form, i.e. it satisfies (\ref{Knoscale})
for $\{ \tilde{\imath} \}=\{U_\lambda \}$.

\subsubsection{No-scale quaternionic breaking} \label{IIAquat}

The no-scale structure of the last piece of (\ref{kahlerlarge})
tells us that if the superpotential does not depend on the complex structure deformations,
 i.e. $\partial_{U_\lambda}W=0$, we obtain a no-scale
supersymmetry breaking vacua in the large complex structure limit by
demanding $D_S W=D_{T^a} W =0$. The moduli whose F-terms are
non-zero are the ones in $\Pi$, which descend from $\N
=2$ hypermultiplets spanning a quaternionic manifold. Therefore
this case belongs to the same class of quaternionic breaking solutions
discussed in the previous section, for which $G$ is an ISD (poly)-form.
In order to make this statement more precise, let us compute the F-terms corresponding to
(\ref{superIIA}). These result in\footnote{The derivation of (\ref{FIIA})
deserves some explanation. First, in order to take the derivative with respect to the
K\"ahler moduli $T^a$, defined in (\ref{Ta2}), we have reexpressed $G$ as,
\begin{equation}
G=F+iC\RE d_H\Omega = e^B[\bar F + d_{\bar H}(e^{-B}\Pi)] \ \nn ,
\end{equation}
with $\Pi$ given in (\ref{pipri}). Since $\langle e^{-B} \Phi , e^{-B} \Psi \rangle=\langle \Phi ,  \Psi \rangle$ for any $\Phi$ and $\Psi$, we
can freely wedge both sides
of the Mukai pairing in (\ref{superIIA}) by $e^{-B}$. Moreover, to
get the last equality we have expressed also
\begin{equation}
D_{T^a} W= -i\int \langle \omega_{a} e^{-iJ}, G \rangle
+i\frac{\mathcal{N}_{\Omega}}{{\cal N}_J} \int \langle \omega_a
e^{-iJ}, e^{iJ} \rangle G^+_{(1)}\ \nn .
\end{equation}
Using (\ref{gammas}) then we can write
\begin{equation}
\omega_a \Phi^+ = \frac{1}{2} (\omega_a)_{mn} dx^m \wedge dx^n
\Phi^+= \frac{1}{4} [ \gamma^m, \{\gamma^n , \Phi^+ \} ]\ \nn ,
\end{equation}
and finally, using  the bispinor expression for $e^{-iJ}$ given in (\ref{SU3}),
the relation between the Mukai pairing and the norm of bispinors
(\ref{mukaispi}), the decomposition (\ref{Gplus}) and the bilinears (\ref{bilinears}),
we arrive to the expression (\ref{FIIA}).}
\begin{align} \label{FIIA} &D_{T^a} W=
\frac{1}{\mathcal{N}_J}\int \left(\mathcal{N}_J\langle
\partial_{T^a} e^{-iJ}, G \rangle - {\cal N}_{J,T^a}\langle e^{-iJ},
G \rangle \right) = -4 {\cal N}_J G^-_{mn} J^{mp} J^{qn} (\omega_a)_{pq} \ , \\
& D_S W= \int \langle e^{-iJ}, G^* \rangle = -8 i {\cal N}_{\Omega} (G^-_{(1)})^* \ . \\
& D_{U_\lambda} W=  W\partial_{U_\lambda} K \ , \label{FIIAend}
\end{align}
where we have already imposed $\partial_{U_\lambda} W=0$ to
compute $D_S W$. We therefore get a no-scale vacua if the
following conditions are satisfied \beq G^-_{mn}=0 \ , \qquad
G^-_{(1)}=0 \ , \qquad \int H\wedge \beta^\lambda=\int dJ\wedge
\beta^\lambda = 0 \ , \label{nsIIA}\eeq where the last two are
required to get $\partial_{U_\lambda}W=0$. These conditions imply
that all NS fluxes ($H_3$ plus torsion) are determined in terms of
the dilaton, the RR singlet fluxes and $F_2^{(8)}$ (see definitions in
Appendix \ref{su3decomp}),
\begin{align}
&\RE {\cal W}_1=\frac{1}{6} e^{\phi^{(4)}} {\cal N}_{\Omega}^{1/2} F_2^{(1)}\ ,  &  &H^{(1)}=
\frac{1}{3}\frac{{\cal N}_J}{ {\cal N}_{\Omega}} F_0\ ,  &
 &F_4^{(1)}=F_6^{(1)}=0\ , \nn \\
&\IM {\cal W}_2=0 \ , &  &\RE {\cal W}_2 \wedge J= -e^{\phi^{(4)}} {\cal N}_{\Omega}^{1/2} *F_2^{(8)}\ ,  &  &F_4^{(8)}=0\ , \nn \\
 &H^\lambda=-\frac{1}{2} \frac{{\cal N}_J}{ {\cal N}_{\Omega}^{1/2}} e^{\phi^{(4)}} F_0 \, \IM Z^{\lambda}\ ,  &  &H_0 = \frac{1}{\RE Z^0} H^\lambda \, \RE({\cal F}_{\lambda})\ , & & \nn \\
  &{\cal W}_3^\lambda=-\frac{1}{4}  \frac{{\cal N}_J}{ {\cal N}_{\Omega}} F_2^{(1)} \, \IM Z^{\lambda}\ ,
 & &({\cal W}_3)_0 = \frac{1}{\RE Z^0} {\cal
W}_3^\lambda \, \RE({\cal F}_{\lambda})\ , \label{su3iiasol} & &
\end{align}
where we have expanded $H^{(6)}=H^\lambda \alpha_\lambda + H_0
\beta^0$, and similarly for ${\cal W}_3$. Notice that the
difference with respect to the supersymmetric solution is
precisely the singlets (while the ${\bf 8}$ component has the same
form as the supersymmetric one, analogously to the ${\bf 6}$ in
type IIB). Moreover, following the same arguments as in type IIB, we expect the warp factor to
behave as in the supersymmetric solution, namely
 \begin{equation}
2i\mathcal{W}_5^*=-e^{\phi}F_2^{(3)}=2i\bar{\partial}A=\frac{2}{3} i\overline{\partial}\phi\
, \label{warprelO6}
\end{equation}
and ${\cal W}_4=0$, so from (\ref{su3iiasol}) and (\ref{warprelO6}) we see that $G$ is
indeed an ISD form.

Inspired by the type IIB no-scale solutions with O5-planes of
previous sections, we may also consider a slightly different class of solutions,
on which $K_{U_\lambda}$ splits as
\begin{equation}
K_{U_\lambda}=-\log(U+ U^*)+K'_{U_{\tilde \lambda}} \quad \
, \quad K'^{\tilde \lambda \bar{\tilde \rho}} K'_{\tilde \lambda} K'_{\bar{\tilde \rho}}=2 \ .
\label{suu}
\end{equation}
This will be the case for example for twisted tori, where $U$ is made out of the
real complex structure of one of the $T^2$ and its axion partner. A no scale solution
arises if
$\partial_{S}W=\partial_{U_{\tilde \lambda}}W=0$, for
$U_{\tilde \lambda}\neq U$, and $D_{T^a}W=D_{U}W=0$. Notice
that the F-terms in this case read,
\begin{equation}
D_{U} W= \int \langle e^{-iJ}, G^* \rangle = -8 i {\cal
N}_{\Omega} (G^-_{(1)})^* \ , \quad D_{S} W=
W\partial_{S}K \ , \quad D_{U_{\tilde \lambda}} W=
W\partial_{U_{\tilde \lambda}} K \ \ (U_{\tilde \lambda}\neq  U) \ ,
\end{equation}
with $D_{T^a}W$ still given by the first line of (\ref{FIIA}).
Thus, we get back again the conditions (\ref{nsIIA}), with
$\lambda$ now running over all  $U_{\tilde \lambda}$, and
$H \wedge \alpha_0=dJ \wedge \alpha_0 = 0$. The breaking is again
mediated by the $\mathcal{N}=1$ scalars descending from the $\mathcal{N}=2$
hypermultiplets, the only difference being the particular directions of the
quaternionic space which enter the breaking.

\subsubsection{No-scale mixed breaking: Scherk-Schwarz breaking} \label{IIAmixed}

Apart from the no-scale solutions with the supersymmetry
spontaneously broken by moduli in $\Pi$ (complex structure and dilaton),
in geometric type IIA compactifications with $SU(3)$ structure there is
another class of solutions on which the breaking involves also
F-terms associated to the moduli in $\Phi_1=\Phi_+$, descending from
$\N=2$ vector multiplets. These solutions are
therefore not dual to the quaternionic breaking solutions, and we believe their type IIB
counterparts correspond to non-geometric compactifications. The
existence of these solutions was noticed in \cite{cfi} from the
four dimensional point of view, however the ten dimensional
construction was missing. Here, we will show that they are related
to non-supersymmetric Scherk-Schwarz compactifications~\cite{schsch}.

Indeed, consider the internal manifold to be a trivial $T^2$
fibration over a base ${\cal B}$, i.e. ${\cal M}=T^2\times \mathcal{B}$, so that the K\"ahler form is
decomposed as $J=J_{T^2}+J_{\mathcal{B}}$, with  $dJ_{T^2}=0$,
 and
$K_{U_{\lambda}}$ satisfying (\ref{suu}). Let us take vanishing fluxes,
$F$=$H$=0, and therefore the setup is of Scherk-Schwarz type. The superpotential becomes
\begin{equation}
W=\int \langle e^{-iJ_{\mathcal{B}}}, iC\RE d\Omega \rangle \ .
\end{equation}
Since it is independent of the
K\"ahler modulus of the $T^2$ fibration, $\tilde T$, it leads to
a no-scale structure when $ J_{\mathcal{B}} \wedge
\partial_{U_\lambda}(d\Omega)=0$, for $U_\lambda \neq \tilde U$.
Notice also that $\RE G=0$, so $G$ is a pure imaginary (poly)-form
and the IASD $SU(3)$ components are therefore automatically determined by the ISD components,
$G^+_{(1)}=G^-_{(1)}$ and $G^+_{mn}=(G^-_{mn})^T$.

Computing the F-terms,
\begin{align}
D_SW&=\frac{i}{2}\int\langle e^{-iJ_{\mathcal{B}}}, d(\alpha_0+\frac{ \tilde u}{s}\beta_{\tilde{U}}) \rangle\ , \\
D_{T^a}W&=\begin{cases}W\partial_{\tilde T}K &
\textrm{for }T^a=\tilde T \\
32 i{\cal N}_J G^-_{mn} J_{\mathcal{B}}^{mp} J_{\mathcal{B}}^{qn}
(\omega_a)_{pq}& \textrm{for }T^a\neq\tilde T
\end{cases}\ , \\
D_{U_\lambda}W&=\begin{cases}W\partial_{U_\lambda}K & \textrm{for
}U_\lambda\neq \tilde U\\
-\frac{s}{\tilde u}D_SW & \textrm{for } U_\lambda = \tilde U
\end{cases}\ ,
\end{align}
we get that in order to have $D_{T^a}W=D_S W=0$, for
$T^a\neq\tilde T$, $G_{mn}^\pm$ has to vanish along the directions
of $\mathcal{B}$ and,
\begin{equation}
\int J_{\mathcal{B}}\wedge d(\alpha_0+\frac{\tilde
u}{s}\beta_{\tilde{U}})=0 \ . \label{sscond}
\end{equation}
Notice that in some sense $S$ and $\tilde{U}$ behave as a single
modulus of the compactification. This will be made more explicit
in a concrete example in next section.

\subsection{Examples} \label{IIAnoscaleex}

\subsubsection{No-scale quaternionic breaking} \label{IIAnoscalequatex}

We consider here a representative of the first class of no-scale
vacua discussed above, i.e. those on which the supersymmetry is
spontaneously broken by F-terms associated exclusively to the
$\mathcal{N}=1$ fields descending from $\mathcal{N}=2$
hypermultiplets ($S$ and $U_\lambda$). These IIA solutions
are mirror to the usual no-scale solutions of type
IIB with O3-planes, or T-dual to the ones with O5-planes discussed
in previous sections. This particular example corresponds to the
ten dimensional realization of one of the no-scale vacua
considered in \cite{cfi}.

We take the internal manifold to be a compact $S^1$ fibration over $T^5$, with O6-planes
wrapping $x^1, x^2, x^3$ and
\begin{equation}
de^1=de^2=de^4=de^5=de^6=0\quad
, \quad de^3=-e^4\wedge e^5 \ .
\end{equation}
To avoid cluttering, let us first give the solution in the limit $A\to 0$ and then comment on how to introduce the warp factor.
Choosing,\footnote{In terms of ${\cal N}=1$ moduli, $C \RE \Omega=
\frac{e^{-\phi_{(4)}}}{\sqrt{\tau^1 \tau^2 \tau^3}} (e^1
\wedge e^2 \wedge e^3 -\tau^{1} \tau^2 e^4
\wedge e^5 \wedge e^3 -\tau^{1} \tau^3 e^4
\wedge e^2 \wedge e^6 -\tau^{2} \tau^3 e^1
\wedge e^5 \wedge e^6) \equiv s e^1
\wedge e^2 \wedge e^3 - u_3   e^4 \wedge
e^5 \wedge e^3 - u_2 e^4 \wedge e^2
\wedge e^6 - u_1 e^1 \wedge e^5 \wedge e^6$.}
\begin{equation} \label{JOIIAquat}
\Omega = (e^1+i\tau^1e^4)\wedge
(e^2+i\tau^2e^5)\wedge
(e^3+i\tau^3e^6)\ , \quad J=-t_1e^1\wedge
e^4-t_2e^2\wedge e^5-t_3e^3\wedge e^6
\ ,
\end{equation}
we get, $d(\RE \Omega)=e^1\wedge e^4\wedge
e^2\wedge e^5$. The torsion classes are
\begin{align} \label{WexIIA}
&\mathcal{W}_1=\frac{1}{6t_1t_2}\quad , \quad
\mathcal{W}_2=\frac{1}{3t_1t_2}\left(J+i\frac{3t_3}{2\tau^3}z^3\wedge
\bar{z}^3\right)\ , \nn \\
&\mathcal{W}_3=-\frac{it_3 }{8\tau^1\tau^2\tau^3}(z^1\wedge
z^2\wedge \bar{z}^3+z^1\wedge \bar{z}^2\wedge
z^3+\bar{z}^1\wedge z^2\wedge z^3)\ +\ \textrm{c.c.}
\end{align}
where $z^a=e^a+ i \tau^a e^{a+3}$.
On top of this, we parameterize a possible expectation value of
the NSNS 3-form as,
\begin{equation} \label{HexIIA}
H=m\frac{t_1t_2t_3}{s}e^4\wedge e^5\wedge
e^6 \ .
\end{equation}
This, together with the ISD condition, determines $G$ as,
\begin{equation} \label{GexIIA}
G=-m-\frac{st_3}{t_1t_2}e^3\wedge
e^6+is e^1\wedge e^4\wedge
e^2\wedge
e^5+imt_1t_2t_3e^1\wedge e^2\wedge e^3\wedge e^4\wedge e^5\wedge e^6\
,
\end{equation}
from which we read the expectation values of the RR field
strengths,
\begin{equation}
F_0=-m \ , \quad F_2=-\frac{st_3}{t_1t_2}e^3\wedge
e^6\ , \quad F_4=F_6=0\ ,
\end{equation}
and the Bianchi identity,
\begin{equation} \label{BexIIA}
dF_2=\frac{st_3}{t_1t_2}e^4\wedge e^5\wedge
e^6=\delta_{D6/O6}\ .
\end{equation}

In terms of $SU(3)$ representations, the only non-vanishing
components of $G$ are,
\begin{equation}
G^+_{(1)}=-\frac{t_3(mt_1t_2+is)}{4\tau^1\tau^2\tau^3}\ , \qquad
G^+_{m\bar n}=\frac{1}{4}\frac{\tau^1\tau^2\tau^3}{t_1t_2t_3}\begin{pmatrix}(G_{(1)}^+)^*\frac{t_1}{\tau^1}&0&0\\
0&(G_{(1)}^+)^*\frac{t_2}{\tau^2}&0\\
0&0&G_{(1)}^+\frac{t_3}{\tau^3}
\end{pmatrix}
\end{equation}
Notice in particular the independence of
$\mathcal{N}_{\Omega}G^+_{(1)}$ (i.e., of the superpotential) on the complex structure moduli,
accordingly with the no-scale structure.

One can make contact with the results of \cite{cfi} by decomposing
the field-strengths between the Chern-Simons couplings and the
background field. Indeed, from (\ref{fstr}) we see that the
VEV's for the axionic parts of the K\"ahler moduli and $S$ are
fixed as,
\begin{equation}
\IM T_a=\frac{1}{m}\int (F_2-\overline{F}_2)\wedge
\tilde{\omega}_a\ , \quad \IM S=\int
(\overline{F}_4+\frac{1}{2m}\overline{F}_2\wedge
\overline{F}_2)\wedge e^3\wedge e^6\ ,
\end{equation}
in agreement with the results of \cite{cfi}.

As argued in the previous sections, the warp factor behaves like in the supersymmetric case,
i.e. the warped solution is obtained by making the replacement $e^a \to e^{A} e^a$,
$e^{a+3} \to e^{-A} e^{a+3}$, $a=1,2,3$, in (\ref{JOIIAquat}). The torsion classes
${\cal W}_1$, ${\cal W}_2$ and $\cW_3$ in (\ref{WexIIA}), as well as $H$ in (\ref{HexIIA}) get multiplied by a factor of $e^{3A}$. Additionally, ${\cal W}_2$ gets a term of the form $-2i \epsilon_{ijk} \frac{\tau^i}{t_i}
\del_{\ib}A \, \bar z^j \wedge z^k$, and $F_2$ a term $s\,  e^{A} *_3 d(e^{-4A})$ (where $*_3$ is the
Hodge dual on the 3-dimensional subspace 456). Besides, $\del \phi= 3 \cW_5= 3 \del A$, as expected from
(\ref{warprelO6}) while $\cW_4$ is zero. Finally, the Bianchi identity
(\ref{BexIIA})  gets an additional term $-2 s e^{-2A} \nabla^2(e^{-4A}) e^4 \wedge e^5 \wedge e^6$.

\subsubsection{No scale mixed breaking} \label{IIAnoscalemixedex}

Here we consider a representative example of this class of
solutions, based on an algebraic solvmanifold with,
\begin{align}
&de^1=de^4=0\nn \ , \\
&de^2=e^6\wedge e^4 \quad , \quad
de^5=e^3\wedge e^4 \ , \nn \\
&de^3=e^4\wedge e^5 \quad , \quad
de^6=e^4\wedge e^2 \ , \label{mixedstr}
\end{align}
As shown in \cite{scan}, this solvmanifold admits a flat metric\footnote{Changing the sign
of $f^3_{45}$ and $f^5_{34}$, it admits also supersymmetric backgrounds
without flux.},
a lattice $\Gamma$ such that the quotient $G/\Gamma$ is compact,
and O6-planes spanning the directions 123, 156, 426, 453, 125
and/or 136. The moduli
space is composed of two K\"ahler moduli, $T_1$ and $T_2$, two
complex structure moduli, $U_2$ and $U_3$, and a single
axio-dilaton $S$.\footnote{The solution requires
$u_1 = s$ (i.e. $\tau_2\tau_3=1$) and
$t_2=t_3$. The first condition guarantees that
(\ref{sscond}) is satisfied, whereas the second one implies that
$G_{mn}^\pm$ takes non-zero values only along the directions of
$e^1$ and $e^4$.} In terms of these, $J$ and $\RE
\Omega$ read (again in the limit $A \to 0$)
\begin{eqnarray}
J&=&-t_1 e^1\wedge e^4\ -\ t_2( e^2\wedge
e^5+ e^3\wedge e^6) \ , \nn \\
 \RE \Omega&=& e^1\wedge
e^2 \wedge e^3 \ - e^1\wedge
e^5 \wedge e^6 -\ e^4\wedge
\left(\frac{u_2}{s}\ e^2\wedge e^6+\frac{u_3}{s}\ e^5\wedge e^3\right)\ ,
\end{eqnarray}
and hence, for $H$ and all the RR forms vanishing, $G$ is given
by
\begin{equation}
G=-2i s e^1\wedge
e^4\wedge(e^2\wedge e^5+ e^3\wedge e^6)\
.
\end{equation}
In terms of $SU(3)$ components,
\begin{equation}
G_{(1)}^+=G_{(1)}^-=\frac{is t_2}{2\tau}\ , \qquad
G_{m\bar n}^+=G_{\bar mn}^-=-\frac{s}{8t_2\tau}\begin{pmatrix}i
& 0 & 0 \\
0 & 0 & 0 \\
0 & 0 & 0
\end{pmatrix}\ ,
\end{equation}
with $\tau=\sqrt{u_2 u_3}/s$ the complex structure parameter of the 2-torus spanned
by $e^1$ and $e^4$. Notice that
$\mathcal{N}_{\Omega}G_{(1)}$ is independent of $U_2$, $U_3$ and
$T_1$. On the other hand, the $F$-terms associated to $S$ and
$T_2$ automatically vanish, thus leading to a no-scale structure
with the corresponding axions stabilized as,
\begin{equation}
\IM S = \int \overline{F}_4\wedge (e^2\wedge
e^5+e^3\wedge e^6) \ , \quad \IM T_2= \int
\overline{H}\wedge e^4\wedge (e^5\wedge
e^6-e^2\wedge e^3)\ .
\end{equation}

The torsion classes are
\begin{align}
&\mathcal{W}_1=-\frac{2}{3t_1t_2}\ , \quad
\mathcal{W}_2=-\frac{2}{3t_1t_2}[2t_1 e^1\wedge
e^4-t_2(
e^2\wedge e^5+ e^3\wedge e^6)] \ , \nn \\
&\mathcal{W}_3=\frac{i}{2\tau}t_2\bar{z}^1\wedge z^2\wedge z^3 \ +
\ \textrm{c.c.}
\end{align}

Note that this solution does not have RR flux or $H$. Therefore, the orientifold planes
are not needed to cancel tadpoles. However, without the orientifold projection the moduli
would be those of $\N=2$. We expect this background to be a no-scale supersymmetry breaking
solution also without the orientifolds, since the equations of motion should not be sensible to
the projection. In any case, if there are orientifold planes and consequently D6-branes to cancel
the tadpoles, but such that these are not on top of each other, the warp factor should
behave as in the previous example.

\subsection{Geometrically induced $\mu$-terms on twisted tori in IIA} \label{muIIA}

In type IIA on SU(3) structure manifolds, $\Phi_{1,2}$ are given
respectively by $\Phi_{+,-}$ in (\ref{SU3}). $\theta_+=\theta_- -
\pi/2$, where the phase $\theta_+$ is a choice, and determines the
location of the O6-planes. Choosing $\theta_-=\pi/2$, the
orientifold projection acts as $\sigma (\Omega)=\bar \Omega$. Let
us use real 1-forms $X^i, Y^{\hat{\imath}}$, $i, \hat{\imath} =
1,2,3$, where the orientifold projection acts as
$\sigma(X^i,Y^{\hi})=(X^i,-Y^{\hi})$. The complex 1-forms $Z^i$
and the symplectic form are given by \beq Z^i=X^i + i \,
\tau^i{}_{\hat{\jmath}} Y^{\hat{\jmath}} \ , \qquad J_c = B+i J=
-i \, T_{i \hat{\jmath}} \, X^i \wedge Y^{\hat{\jmath}} \eeq where
$\tau^i{}_{\hat{\jmath}}$ are real, and $T_{i \hat{\jmath}}$ are
complex K\"ahler moduli. These define an SU(3) structure if the
matrix $T \tau^{-1}$ is symmetric, i.e $T_{i \hat{\jmath}}
(\tau^{-1})^{\hat{\jmath}} {}_k= T_{k \hat{\jmath}}
(\tau^{-1})^{\hat{\jmath}}{}_i$. In that case, $B+i J$ is (1,1)
with respect to the complex structure. A basis of 3-forms is given
by
\begin{align}
\alpha_0 = X^1 \wedge X^2 \wedge X^3 \ , \quad  \qquad \ \ & \beta^0 = Y^1 \wedge Y^2 \wedge Y^3 \ . \nn\\
  \alpha_{j}{}^{\hat{\imath}} = \frac{1}{2} \epsilon_{jkl}
X^k \wedge X^l \wedge Y^{\hat{\imath}} \ , \qquad &\beta^{i}{}_{\hat{\jmath}} = -\frac{1}{2} \epsilon_{\hat{\jmath} \hat k \hat l}
Y^{\hat k} \wedge Y^{\hat l} \wedge X^i \ .
\end{align}
The holomorphic 3-form $\Ox$ is given in this basis by \beq
\Omega=\alpha_0 + i \alpha_{j}{}^{\hat{\imath}}
\tau^{j}{}_{\hat{\imath}}  + \beta^{i}{}_{\hat{\jmath}} (\cof
\tau)_{i}{}^{\hat{\jmath}} - i \beta^0(\det \tau) \ , \eeq where
\beq (\cof \tau)_i{}^{\hat{\jmath}} =  (\det \tau) \tau^{-1,{\rm
T}} = \frac{1}{2} \, \epsilon_{ikm} \epsilon^{\hat{\jmath} \hat p
\hat q} \tau^{k}{}_{\hat p} \tau^{m}{}_{\hat q} \ . \eeq A
supersymmetric D6-brane has to satisfy the  D-flatness condition
(\ref{Dflat}), which reads in this case \beq P_\Sigma (\IM \Omega)
= 0 \ . \eeq The cycles that satisfy it are $\Sigma^0$,
$\Sigma^{\hj}{}_{i}$, dual respectively to the even left invariant
forms $\alpha_0$ and $\beta^i{}_{\hat{\jmath}}$. The F-flatness
condition (\ref{Fflat}) implies \beq P_\Sigma[J] =0 \ , \qquad
{\cal F}=0 \ , \eeq i.e. the branes wrap special Lagrangian
submanifolds. The first condition is satisfied automatically on
the cycle $\Sigma^0$, while for the cycles $\Sigma_i{}^{\hj}$ they
impose $T_{i\hat k}=T_{i \hat l}=0$, where ${\hat k}, {\hat l}
\neq \hj$. This means that for a given $i$, there's only one $\hj$
such that   $\Sigma_i{}^{\hj}$ is supersymmetric, namely $\hj$ is
defined by the combination $T_{i \hat m} y^{\hat m}$. There are
therefore in total four supersymmetric cycles, $\Sigma^0$ and
$\Sigma^i$.

The superpotential (\ref{supgen}) for a D6-brane wrapping
$(\Sigma,{\cal F})$ is
 \beq \label{WA}
 W= \frac{1}{4} \int_{\cB_i} e^{3A-\phi} (\tilde F-J_c)^2
 \eeq

For the cycle $\Sigma^0$, dual to $\alpha_0$,
  $\cB_i$ are chains
 dual to the forms $X^1 \wedge X^2 \wedge X^3 \wedge Y^{\hj}$. The holomorphic
 brane fields are given in (\ref{Afields}), and their superpotential
 is
 \beq \label{finmud60}
 W_{D6_0}=\frac{1}{4} \epsilon^{ikl} (T_{k \hat r}  f^{\hat r}_{\hi l}+ T_{r \hi} f^r_{lk} ) T_{i \hj} \phi^{\hi} \phi^{\hj} \ , \qquad \phi_i=A_i - i \, T_{i \hat{\jmath}} y^{\hat{\jmath}} \equiv T_{i \hj} \phi^{\hj}
 \eeq

For the cycle $\Sigma_i$, dual to $\beta^{i}{}_{\hat{\jmath}} = -\frac{1}{2}\,  \epsilon_{\hat{\jmath} \hat k \hat l}
Y^{\hat k} \wedge Y^{\hat l} \wedge X^i$   the superpotential is
\begin{align} \label{finmud6i}
W_{D6_{\textrm i}}=&-\frac{1}{2} (T_{p \hat l} f^p_{\hat k \hj} + 2 T_{p (\hat k} f^p_{\hj) \hat l} ) T_{i \hj}
\, \phi^{\hj} \phi^{\hj} - \frac12 (T_{p \hat k} f^p_{ib} + T_{(i| \hat r} f^{\hat r}_{b)\hat k} ) T_{a \hat l} \phi^b \phi^a \nn \\
&+ \frac12 \left( - (T_{a \hat r} f^{\hat r}_{\hat l \hat k}   + T_{p \hat l} f^p_{\hat k a} ) T_{i \hj}
+ (T_{i \hat r} f^{\hat r}_{\hj \hat k}   + T_{p (\hj} f^p_{\hat k)i}) T_{a \hat l}
\right) \phi^{\hj} \phi^a    \ ,
\end{align}
where
\beq
  \phi_i=A_i - i \, T_{i \hj} y^{\hj }\equiv T_{i \hj} \phi^{\hj}  \ ,  \qquad
\phi_{\hat b}=A_{\hat b} - i \, T_{\hat b a} x^a \equiv T_{\hat b a} \phi^{a}
 \ ,
 \eeq
  $a=\{ k, l \}, \hat b=\{\hat k, \hat l\}$ and antisymmetrization in $\hat k, \hat l$
is understood.

Similarly to the type IIB case, (\ref{WA}) contains also terms that are not holomorphic in the
brane moduli. These are proportional to combinations of structure constants that
break supersymmetry. The ${\cal N}=1$ Minkowski vacuum condition, $dJ_c=0$, requires
\beq \label{dJ0}
T_{[k| \hat{r}} f^{\hat{r}}_{l] \hj}-T_{i \hj} f^i_{kl} =0 \ , \quad T_{i [\hj |} f^i_{\hat l] k} - T_{k \hat r} f^{\hat r}_{\hj \hat l} =0 \ ,  \quad T_{i \hj} f^i_{\hat k \hat l}
\epsilon^{\hj \hat k \hat l}=0  \ , \quad T_{i \hj} f^{\hj}_{kl} \epsilon^{ikl} =0 \  \ .
\eeq
Terms that are holomorphic in the brane moduli appear for example with the combination
$T_{[k| \hat{r}} f^{\hat{r}}_{l] \hj}+T_{i \hj} f^i_{kl}$, while the combination with a minus
sign gives rise to non holomorphic terms and is therefore discarded in (\ref{finmud60}).

\TABLE{ \begin{tabular}{c||c|c}
&D6$_{0}$& D6${}_1$\\
\hline $t_1\mu_{11}$& $\frac{1}{2u_1}(T_{2} f^{\hat 2}_{\hat 1 3}-T_{3} f^{\hat 3}_{\hat 1 2}-2T_{1} f^1_{23} ) $ & $\frac{1}{2s} (T_{2} f^2_{\hat 3 \hat 1} + T_{3} f^3_{\hat 1 \hat 2}-T_{1} f^1_{\hat 2 \hat 3}) $ \\
$t_2\mu_{22}$ & $\frac{1}{2u_2} (-T_{1} f^{\hat 1}_{\hat 2
3}+T_{3} f^{\hat 3}_{\hat 2 1}+
2T_{2} f^2_{31} ) $ &  $\frac{-1}{2u_3} (2T_{3} f^3_{12}+ T_{1} f^{\hat 1}_{2 \hat 3} + T_{2} f^{\hat 2}_{1 \hat 3}) $ \\
$t_3\mu_{33}$ & $\frac{1}{2u_3} (T_{1} f^{\hat 1}_{\hat 3 2}-T_{2}
f^{\hat 2}_{\hat 3 1}+2T_{3} f^3_{12} )$ & $\frac{1}{2u_2} (2T_{2}
f^2_{13}+ T_{1} f^{\hat 1}_{3 \hat 2} + T_{3} f^{\hat 3}_{1 \hat
2}) $
\vspace{0.4cm}\\
&D6${}_2$&D6${}_3$\\
\hline $t_1\mu_{11}$& $\frac{1}{2u_3}(2T_{3} f^3_{21}+ T_{1} f^{\hat 1}_{2 \hat 3} + T_{2} f^{\hat 2}_{1\hat 3 }) $ & $\frac{-1}{2u_2} (2T_{2} f^2_{31}+ T_{3} f^{\hat 3}_{1 \hat 2} + T_{1} f^{\hat 1}_{3 \hat 2}) $ \\
$t_2\mu_{22}$ &   $\frac{1}{2s} (T_{1} f^1_{\hat 2 \hat 3}+T_{2} f^2_{\hat 1 \hat 3} + T_{3} f^3_{\hat 1 \hat 2}) $ &  $\frac{1}{2u_1} (2T_{1} f^1_{32}+ T_{3} f^{\hat 3}_{2 \hat 1} + T_{2} f^{\hat 2}_{3 \hat 1}) $ \\
$t_3\mu_{33}$ & $\frac{-1}{2u_1} (2T_{1} f^1_{23}+ T_{2} f^{\hat 2}_{3 \hat 1} + T_{3} f^{\hat 3}_{2 \hat 1}) $ & $\frac{1}{2s} (T_{1} f^1_{\hat 2 \hat 3}-T_{2} f^2_{\hat 1 \hat 3} - T_{3} f^3_{\hat 1 \hat 2})
 $
\end{tabular}

\caption{Supersymmetric torsion induced $\mu$-terms for D6$_0$ and
D6${}_i$ branes wrapping the cycles dual to $X^1 X^2 X^3$ and
$\epsilon_{ijk} X^i Y^{\hj} Y^{\hat k}$ in a factorizable twisted
torus.}\label{mususyIIA} }

We summarize in table \ref{mususyIIA} the torsion induced
$\mu$-terms for D6-branes on a factorizable torus, where the
structure constants have one leg on each 2-torus. We have set the
normalization of the matter fields to match the result for the
D7-brane effective $\mu$-term~\cite{hepth0408036,0501139} in vacua where a
T-dual description is available.

\section{Soft-terms on twisted tori} \label{soft}

A background where supersymmetry is broken spontaneously by
torsion or fluxes, induces soft-supersymmetry breaking terms on a
D-brane living in it. In this section, we compute the pattern of
soft-terms for factorizable twisted tori in the  no-scale
supersymmetry breaking va\-cua of sections \ref{noscaleIIB} and
\ref{noscaleIIA}.

Bulk and brane sectors combine in an $\N=1$ supergravity. Brane moduli
$\phi^i$ form $\N=1$ chiral superfields charged under a non-Abelian gauge group
(or just a U(1), for a single brane, which will be mostly the case for us).
These couple to the neutral bulk moduli, such that brane
fluctuations enter in the definition of the moduli
descending from $\N=2$ hypermultiplets. The K\"ahler potential is
still given by (\ref{Kgen}), but $\phi^i$ enter in $\Pi$
(\ref{Pi}), and therefore modify the definition of the moduli
(\ref{Ta}), (\ref{spinorexpan}).  These can be found in
\cite{GGJL} and \cite{JL} respectively for D3-branes and D7-branes
in SU(3) structure manifolds, while for the simplest case of
factorizable toroidal models they are given in
\cite{hepph9812397,lustmetrics}. If the gauge symmetry on the
branes is unbroken, the vacuum expectation value of the fields
$\phi^i$ vanishes and it is convenient to expand the K\"ahler
potential in power series of $\phi^i$
\begin{multline} \label{Kexp}
K(M,\bar M, \phi,\bar\phi) = \hat K(M,\bar M) + Z_{i\bar j}(M,\bar M)\, \phi^i\bar\phi^{\bar j}
+ \frac12\big( H_{ij}(M,\bar M)\, \phi^i\phi^{j}
+ \textrm{h.c.} \big)
+ \ldots\ , \\
\equiv \hat K(M,\bar M) + K_{Dp}(M, \bar M, \phi,\bar \phi) \nn
\end{multline}
where $M$ denotes collectively bulk moduli. The function $H_{ij}$
has been found to be nonzero for D3-branes \cite{GGJL} and for
D7-branes in compactifications with an uplift to F-theory
\cite{fkahler}. In the latter case they turn out to be equal to the
$Z_{i\bj}$ terms, $H_{i j}=Z_{i\bj}$. The reason is that the
D-brane moduli enter the K\"ahler potential through terms of the
form $(\phi^i+\bar \phi^i)(\phi^j+\bar \phi^j)$.

Bulk and brane superpotential are also combined in the expansion
\beq \label{Wexp} W(M,\phi) = \hat W(M) + W_{Dp} (M, \phi) =\hat
W(M) + \frac 12\, \tilde \mu_{ij}(M)\, \phi^i\phi^{j} + \frac 16\,
\tilde Y_{ijk} (M)\, \phi^i\phi^{j}\phi^{k} +\ldots\ . \eeq The
gauge couplings obey \beq g_{Dp}^{-2}=2\ \RE f_{Dp}(M) \eeq where $f_{Dp}$ is the
holomorphic gauge kinetic function. Inserting these in
(\ref{Vgen}) and keeping terms up to cubic order in $\phi$, we get
an effective potential for the brane fields in the flat limit
$M_{Pl}\to \infty$, $m_{3/2}$ fixed, of the form \beq\label{Veff}
V^{\eff} = Z^{i\bj}\, (\partial_i \Weff)(\partial_{\jb} \Weffb) +\
m^2_{i\bj,{\rm soft}}\, \phi^i\bar\phi^{\jb} +\frac16\,
A_{ijk}\phi^i\phi^{j} \phi^k + \frac12\, B_{ij}\phi^i\phi^{j} +
\textrm{h.c.}\ , \eeq where \cite{IL,KL,hepph9707209}
\bea\label{susy} \Weff &=&\frac{1}{2}\, \mu_{ij}\, \phi^i \phi^j
+\frac{1}{3}\, Y_{ijk}\, \phi^i \phi^j \phi^k \ ,\\
\mu_{ij}&=& e^{\hat{K}/(2M_{Pl}^2)} \tilde{\mu}_{ij} + m_{3/2} H_{ij} - F^{\bar I} \bar{\partial}_{\bar I} H_{ij}\ ,\nonumber \\
Y_{ijk}&=& e^{\hat{K}/(2M_{Pl}^2)} \tilde{Y}_{ijk}\ .\nonumber \eea
and the soft supersymmetry breaking terms read \bea \label{softi}
m^2_{i \bj,{\rm soft}}&=& |m_{3/2}|^2 \, Z_{i \bj} -
F^I F^{\bar J} R_{I {\bar J} i \bj}  ,\nonumber \\
A_{ijk} &=& F^I D_I Y_{ijk}\ , \\
B_{ij} &=& 2|m_{3/2}|^2 \, H_{ij}
- \bar m_{3/2} \bar F^{\bar J}\bar\partial_{\bar J} H_{ij}
+  m_{3/2} F^{I} D_I H_{ij} \nn \\
&& - F^{I}\bar F^{\bar J}D_I\bar\partial_{\bar J} H_{ij} -
e^{K/(2M_{Pl}^2)} \tilde \mu_{ij}  \bar m_{3/2}
+e^{K/(2M_{Pl}^2)}F^{I}D_I\tilde \mu_{ij} \ ,\nonumber \eea where
\bea\label{curvi} m_{3/2}&=& e^{\hat K/(2M_{Pl}^2)} \frac{\hat
W}{M_{Pl}^2} \ , \qquad \qquad \quad
F^{\bar I} = e^{\hat{K}/(2M_{Pl}^2)} \hat K^{\bar I J} D_{J} \hat W \ ,\nn \\
R_{I \bar J i \bj} &=& \partial_I \bar \partial_{\bar J} Z_{i \bj} -
   \Gamma^{k}_{Ii}Z_{k \bar l}\Gamma^{\bar l}_{\bar J \bj} \ ,\qquad
\Gamma ^l_{Ii}=Z ^{l\bj} \partial_I Z_{\bj i}\ ,\nonumber\\
D_I Y_{ijk} &=& \partial_I Y_{ijk} + \frac{1}{2M_{Pl}^2} \hat{K}_I
Y_{ijk} -
3\Gamma ^l_{I(i} Y_{jk)l} \,  ,\\
D_I  \tilde \mu_{ij} &=& \partial_I \tilde \mu_{ij} +
\frac{1}{2M_{Pl}^2} \hat{K}_I \tilde \mu_{ij} - 2\Gamma ^l_{I(i}
\tilde \mu_{j)l} \ . \nonumber \eea with $M_{Pl}$ the 4d Planck
mass. In these expressions we have taken the bulk moduli to be
dimensionless (i.e. the quantum modes). This amounts to factor out
the volume dependence of the 4d dilaton and the K\"ahler moduli
into powers of $M_{Pl}$. The RR and NSNS forms have units of
$\textrm{(length)}^{-1}$, and $\hat W=M_{Pl}^3 \tilde W$, $\hat
K=M_{Pl}^2\tilde K$, with $\tilde W$ and $\tilde K$ respectively
the dimensionless bulk superpotential and K\"ahler potential. In
these units $\tilde W$ is a polynomial in the bulk moduli with
integer coefficients.

Notice that for a D-brane superpotential of the form (\ref{Wexp}),
the first term in (\ref{Veff}) gives a ``supersymmetric" mass
term, as well as a trilinear $C$-coupling between two holomorphic
and one antiholomorphic brane fields. These are given by
\begin{equation}
m^2_{i \bj, \rm{susy}}= e^{\hat K/M_{Pl}^2} \tilde \mu_{ik}
\bar{\tilde \mu}_{\bar l \bj} Z^{k \bar l} \ , \quad C_{ij \bar k
, \rm{susy}} =  e^{\hat K/M_{Pl}^2} \tilde Y_{ijl} \bar{\tilde
\mu}_{\bar m \bar k} Z^{l \bar m} \ .
\label{supermasiv}\end{equation} Apart from these, additional mass
terms and trilinear couplings are generated from $H_{ij}$ through
the Giudice-Masiero mechanism~\cite{giudice}
\begin{align}
\mu_{ik,\rm{GM}}&=m_{3/2}H_{ik}-F^{\bar I}\bar{\partial}_{\bar
I}H_{ik}\ , \\
m^2_{i \bj, \rm{GM}}&=(\mu_{ik,\rm{GM}}\bar \mu_{\bar l\bj,\rm{GM}} + \tilde \mu_{ik}\bar \mu_{\bar l\bj,\rm{GM}} + \mu_{ik,\rm{GM}}\bar{\tilde \mu}_{\bar l\bj} )Z^{k\bar l} \ . \nn
\end{align}
The $H_{ij}$ couplings can be thus absorbed into non-holomorphic
contributions in the bulk moduli to the effective superpotential,
$\Weff$, as already exposed in section \ref{muIIB}.

In a generic compactification the total scalar masses for
the brane fields therefore receive three tree-level contributions,
\begin{equation}
m^2_{i \bj}=m^2_{i \bj, \rm{susy}}+m^2_{i \bj, \rm{GM}}+m^2_{i
\bj, \rm{soft}}\ .
\end{equation}
However, the no-scale condition
(\ref{Knoscale}) often induces systematic cancellations which lead
to vanishing $\mu_{i j, \rm{GM}}$ and $m^2_{i \bj, \rm{soft}}$.
More precisely, parameterizing $Z_{i\bj}$ and $H_{ij}$ as
\begin{equation} \label{HZ}
Z_{i\bj}=H_{ij}=\prod_I
\frac{\textrm{const.}}{(M^I+\bar{M}^I)^{\alpha_I}}\ ,
\end{equation}
with $M^I$ the collective bulk moduli, the condition to have
$\mu_{i j, \rm{GM}}=m^2_{i \bj, \rm{soft}}=0$ is given by
\begin{equation}
\sum_{I, \ F^I\neq 0}\alpha_I = 1\ . \label{condmass}
\end{equation}
In that case, it is easy to show additionally that $B_{ij}$ gets
no contribution from $H_{ij}$, i.e. the first four terms in the
expression for $B_{ij}$ in (\ref{softi}) also cancel.

The $\mu$-terms computed in sections \ref{muIIB} and \ref{muIIA}
are the total effective $\mu_{ij}$ of (\ref{susy}). In
vacua where a T-dual description is available, they indeed match
correctly the effective D7-brane $\mu$-term computed in
\cite{hepth0408036} by dimensional reduction of the DBI-CS action, and
in \cite{0501139} by analysis of the effective supergravity
in compactifications of F-theory. Thus, in what follows we
will not consider explicitly the $H_{ij}$
couplings, but instead we will work in terms of $\Weff$.

Before we move on to the specific supersymmetry breaking vacua,
let us remark that all these are contributions coming from pure
moduli mediation. As noticed in \cite{hepth0503216},
non-perturbative or loop contributions such as anomaly mediation
may be generically as important as moduli mediation contributions,
and therefore in a concrete phenomenological model they should be
taken into account.

\subsection{Quaternionic breaking} \label{softquat}

We will compute soft-terms in the case of quaternionic breaking
for D9 and D5-branes. The case of D6-branes in this type of
supersymmetry breaking vacua can be easily obtained by T-duality.
The gauge kinetic couplings and K\"ahler potential $K_{D5}$,
$K_{D9}$ for factorizable toroidal compactifications has been
computed in~\cite{hepph9812397,lustmetrics}, obtaining up to
second order in $\phi$
\begin{equation}
K_{D9}= \sum_i^3\frac{|\phi^i|^2}{(U^i+\bar U^i)(T^i+\bar T^i)} \
, \quad
K_{D5_k}= \sum_{i,j=1}^3d_{ijk}\frac{|\phi^j|^2}{(T^i+\bar T^i)(U^j+\bar U^j)} + \frac{|\phi^k|^2}{(S+\bar S)(U^k+\bar U^k)} \ , \nn
\end{equation}
\begin{equation}
f_{D9}= S \ , \qquad \qquad \qquad f_{D5_k}= T^k \ , \label{KD9D5}
\end{equation}
where the $D5_k$-branes are wrapping the $k$-th 2-torus and
$d_{ijk}=1$ for $i\neq j \neq k$, and $0$ otherwise.

We consider no-scale vacua with supersymmetry spontaneously broken
by the quaternionic sector. For that purpose we take the internal
twisted torus to be a torus fibration over another torus, e.g. of the third torus
over the first and second tori. The fibration is fully
parameterized by the structure constants
$f^{\bar{3}}_{\bar{1}\bar{2}}$, $f^{3}_{\bar{1}2}$,
$f^{3}_{1\bar{2}}$ and $f^{3}_{\bar{1}\bar{2}}$, with all the
other structure constants zero. Assuming $G_{(1)}^+$ is
independent of the K\"ahler modulus of the fiber, $T^3$ (i.e., $D_{T_3} W=0$ and $W$ is independent
of $S,T_1$ and $T_2$), we obtain
the gravitino mass,
\begin{equation}
m_{3/2}=M_{Pl} e^{\hat{K}/(2M_{Pl}^2)}
(U^1+\overline{U}^1)(U^2+\overline{U}^2)(T^3+\overline{T}^3)f^3_{\bar{1}\bar{2}}\
.
\end{equation}
and the scalar potentials for the light scalar modes of the
D-branes in the no-scale vacuum are,\footnote{\label{foot:resc}We
take the usual rescaling of the matter fields, $\phi^i \to
(Z_{i\bi})^{-1/2}\phi^i$, in order to have canonically
normalized kinetic terms.}
\begin{align}
&V_{D9}=\frac{e^{\hat{K}/M_{Pl}^2}}{M_{Pl}^4}\left|M_{Pl}^2\partial_{\phi^3}W_{D9}+(\phi^3)^*  \hat W \right|^2 \ , \label{vvd95} \\
&V_{D5_1}=\frac{e^{\hat{K}/M_{Pl}^2}}{M_{Pl}^4}|M_{Pl}^2\partial_{\phi^3}W_{D5_1}+(\phi^2)^*\hat W |^2 \ , \nn \\
&V_{D5_2}=\frac{e^{\hat{K}/M_{Pl}^2}}{M_{Pl}^4}|M_{Pl}^2\partial_{\phi^1}W_{D5_2}+(\phi^1)^*\hat W |^2 \ , \nn \\
&V_{D5_3}=0 \ , \nn
\end{align}
for pure moduli mediation.

We have summarized in Table \ref{softfin} the pattern of soft
supersymmetry-breaking terms which results of plugging
(\ref{supervac}) and (\ref{finmud9b}) into the above scalar
potentials. For that we have assumed the usual superpotential
trilinear couplings, $\tilde Y_{ijk}=\epsilon_{ijk}$. The rescaled
structure constants, $\tilde f^I_{JK}$, are defined in
(\ref{renorma}). The resulting pattern is clearly related to the
one arising in the worldvolume of D3 and D7-branes in the dual
compactification~\cite{hepth0311241,GGJL,0406092,hepth0408036}.
Indeed, T-dualizing along the third torus, $D5_3$-branes are
mapped to D3-branes, whereas $D5_1$, $D5_2$ and $D9$-branes are
mapped respectively to $D7_2$, $D7_1$ and $D7_3$-branes. As
expected, the light modes of $D5_3$-branes remain massless,
whereas only one complex geometric moduli of the $D5_1$, $D5_2$
and $D9$-branes becomes massive, corresponding to the geometric
moduli of the dual D7-brane. This structure of zero modes can be understood in terms
of the condition (\ref{condmass}). Indeed, making use of (\ref{KD9D5}), we get
that $\mu_{ij,\rm{GM}}=m_{i\bj,\rm{soft}}^2=0$ for all the scalars in
the $D5_3$-branes, and all the scalars but  $\phi^1$, $\phi^2$ and $\phi^3$
in the $D5_2$, $D5_1$ and $D9$-branes respectively, in agreement with the results
of table \ref{softfin}.

\TABLE{
\begin{tabular}{|c||c|c|c|c|}
\hline
& $D9$ & $D5_1$ & $D5_2$ & $D5_3$ \\
\hline
\hline
$\mu_{11}$ & $0$ & $0$ & $4e^{\hat K/2}\tilde f^3_{1\bar{2}}t_3$ & $0$ \\
$\mu_{22}$ & $0$ & $4e^{\hat K/2}\tilde f^3_{\bar{1}2}t_3$ & $0$ & $0$ \\
$\mu_{33}$ & $4e^{\hat K/2}\tilde f^{\bar{3}}_{\bar{1}\bar{2}}t_3$ & $0$ & $0$ & $0$ \\
\hline
$m_{1\bar 1}^2$ & $0$ & $0$ & $|\mu_{33}|^2+|m_{3/2}|^2$ & $0$ \\
$m_{2\bar 2}^2$ & $0$ & $|\mu_{22}|^2+|m_{3/2}|^2$ & $0$ & $0$ \\
$m_{3\bar 3}^2$ & $|\mu_{11}|^2+|m_{3/2}|^2$ & $0$ & $0$ & $0$ \\
\hline
$B_{11}$ & $0$ & $0$ & $2 \mu_{33} \bar m_{3/2}$ & $0$ \\
$B_{22}$ & $0$ & $2 \mu_{22} \bar m_{3/2}$ & $0$ & $0$ \\
$B_{33}$ & $2 \mu_{11} \bar m_{3/2}$ & $0$ & $0$ & $0$ \\
\hline
$A_{123}$ & $g_{D9}m_{3/2}$ & $g_{D5_1}m_{3/2}$ & $g_{D5_2}m_{3/2}$ & $0$ \\
\hline
$C_{1\bar 2\bar 3}$ & $0$ & $0$ & $ \mu_{33} \, g_{D5_2}$ & $0$ \\
$C_{\bar 1 2\bar 3}$ & $0$ & $\mu_{22} \, g_{D5_1}$ & $0$ & $0$ \\
$C_{\bar 1 \bar 2 3}$ & $\mu_{11} \, g_{D9}$ & $0$ & $0$ & $0$ \\
\hline
\end{tabular}
\caption{Torsion induced soft parameters for $D9$, $D5_1$, $D5_2$
and $D5_3$-branes, in a no-scale vacuum of a factorizable twisted
torus with $W$ independent of $S, T_1, T_2$, and $D_M W=0$ for the
remaining moduli. The gauge coupling constants are $g_{D9}=(S+\bar
S)^{-1/2}$ and $g_{D5_k}=(T^k+\bar T^k)^{-1/2}$, and we have set $M_{Pl}=1$.}
\label{softfin} }

Both in the supersymmetric and in the no-scale cases, the
couplings induced by $\mathcal{W}_3$ (i.e., by the structure
constants $f^{\bar{3}}_{\bar{1}\bar{2}}$, $f^{3}_{\bar{1}2}$) give
rise to masses and C-terms satisfying (\ref{supermasiv}) and
therefore are compatible with $\mathcal{N}=1$ supersymmetry with
some massive chiral supermultiplets. On the other hand, $\mathcal{W}_1$,
proportional to $f^3_{\bar{1}\bar{2}}$, gives rise to couplings
satisfying
\begin{equation}
\textrm{Tr}(m_{i, \rm{soft}}^2)=m_{3/2}^2 \quad , \quad A_{ijk}=h_{ijk} \textrm{Tr }(m_{i, \rm{soft}}^2)\ , \label{n4finito}
\end{equation}
with, $h_{ijk}=e^{\hat K/2M_{Pl}}\epsilon_{ijk}(Z_{i\bi}Z_{j\bj}Z_{k\bar k})^{-1/2}$ the physical Yukawa.
This behavior was already observed in
\cite{hepth0311241,GGJL,0406092,hepth0408036} for D3 and D7-branes
in the presence of 3-form fluxes. D6-branes in vacua where
supersymmetry is broken by the quaternionic sector follow the same
pattern of soft supersymmetry breaking terms.

The pattern of moduli mediated soft supersymmetry-breaking terms
therefore can be recast in terms of a small set of parameters: the
gravitino mass plus some topological $\mu$-terms for each stack of
branes. Hence, consider for example the no-scale $K3 \times T^2$
fibration of section \ref{sec:exIIB}. In a complex basis the
structure constants (\ref{tdual2}) read
\begin{equation}
f^3_{\bar 1 2}=f^3_{1\bar 2}=-f^{\bar 3}_{\bar 1 \bar
2}=\frac{1+iu}{4u^2}\ .
\end{equation}
From (\ref{vvd95}) we then obtain the tree-level scalar potentials
for the D-brane fields
\begin{align}
V_{D9}&=\left|\frac{\phi^1\phi^2}{\sqrt{2s}}-\frac{(1+iu)\phi^3+t(3iu+1)(\phi^3)^*}{(32\tilde t_1\tilde t_2 t u^3)^{1/2}}\right|^2\\
V_{D5_1}&=\left|\frac{\phi^1\phi^3}{\sqrt{2\tilde t_1}}-\frac{(1+iu)\phi^2-t(3iu+1)(\phi^2)^*}{(32\tilde t_1\tilde t_2 t u^3)^{1/2}}\right|^2\nn \\
V_{D5_2}&=\left|\frac{\phi^2\phi^3}{\sqrt{2\tilde t_1}}+\frac{(1+iu)\phi^1-t(3iu+1)(\phi^1)^*}{(32\tilde t_1\tilde t_2 t u^3)^{1/2}}\right|^2\nn \\
V_{D5_3}&=0\nn
\end{align}
More detailed phenomenological analysis for this class of vacua in the D3/D7
setup, taking into account other effects such as non-perturbative effects
or $\alpha'$ corrections to the K\"ahler potential, can be found
in \cite{d3d7pheno,d3d7large,hepth0503216,kkltpheno}.

\subsection{Mixed breaking} \label{sofmixed}

For type IIA we have seen in section \ref{IIAmixed} that there is
another class of no-scale vacua where supersymmetry is broken by
moduli belonging in part to descendants of $N=2$ hypermultiplets,
and in part to descendants of scalars in vector multiplets. In
this section we compute the pattern of soft-terms for D6-branes
placed on this type of vacua in factorizable (twisted) $T^6$
models, with 3 complex structure and 3 K\"ahler moduli, and
structure constants with one leg on each torus. We consider a
no-scale vacua where supersymmetry is broken by $T_1, U_2, U_3$,
i.e. $\del_{T_1}W=\del_{U_2} W=\del_{U_3} W=0$ in the vacuum,
while $D_{T_2} W=D_{T_3} W=D_{U_1} W=D_S W=0$. Supersymmetry
breaking is due solely to the torsion in this class of vacua, i.e.
$H$ and all RR fluxes are zero. The bulk superpotential is \beq
\label{Wmixedfacto} \hat W=M_{Pl}^2e^{-\hat K/(2M_{Pl}^2)} m_{3/2} =
M_{Pl}^3\left[S T_2 f^2_{\hat 1 \hat 3} - S T_3 f^3_{\hat 1 \hat
2}+ U_1 T_2 f^{\hat 2}_{\hat 1 3} - U_1 T_3  f^{\hat 3}_{\hat 1
2}\right] \eeq The other structure constants allowed in a
factorizable torus vanish in this type of vacua.

As explicitly shown in the particular example of section \ref{IIAnoscalemixedex}, the imaginary parts
of the K\"ahler moduli appearing in the superpotential, $\textrm{Im }T_{2}$ and $\textrm{Im }T_{3}$,
are related to the background $\overline{H}$. Indeed,
from $H=0$ we get
\begin{equation}
\overline{H}=-dB=-(f^2_{\hat 1\hat 3}\textrm{Im }T_2-f^3_{\hat 1\hat 2}\textrm{Im }T_3) e^4\wedge e^5\wedge e^6
+(f^{\hat 2}_{\hat 1 3}\textrm{Im }T_2-f^{\hat 3}_{\hat 1 2}\textrm{Im }T_3) e^4\wedge e^2\wedge e^3\ .
\end{equation}
Since $\overline{H}\neq 0$ may induce additional $\mu$-terms in the
D6-branes which we are not computing here, in what follows we
set $\overline{H}=\textrm{Im }T_2=\textrm{Im }T_3=0$. The final result however
should not depend on this choice, as the VEV for physical field, $H$,
is fixed.

The K\"ahler potential for D6-branes is the T-dual version of
(\ref{KD9D5}), where we should exchange K\"ahler and complex
structure moduli, $D9$ by $D6_0$ and $D5_k$ by $D6_k$. We get
\begin{equation*}
K_{D6_0}= \sum_i^3\frac{|\phi^i|^2}{(U^i+\bar U^i)(T^i+\bar T^i)}
\ , \qquad K_{D6_k}=
\sum_{i,j=1}^3d_{ijk}\frac{|\phi^j|^2}{(U^i+\bar U^i)(T^j+\bar
T^j)} + \frac{|\phi^k|^2}{(S+\bar S)(T^k+\bar T^k)} \ ,
\end{equation*}
\begin{equation}
 f_{D6_0}= S \ , \qquad \qquad \qquad  f_{D6_k}= U^k \ . \label{KD66}
\end{equation}

Rescaling the matter fields as in the quaternionic breaking (see
footnote \ref{foot:resc}), we get the following potential for
D6-branes in these vacua up to cubic order in $\phi^i$,
\begin{align}
&V_{D6_0}=\frac{e^{\hat{K}/M_{Pl}^2}}{M_{Pl}^4}\left|M_{Pl}^2\partial_{\phi^1}W_{D6_0} \right|^2 \ , \label{d6gg} \\
&V_{D6_1}=\frac{e^{\hat{K}/M_{Pl}^2}}{M_{Pl}^4}\left|M_{Pl}^2\partial_{\phi^1}W_{D6_1} \right|^2 \ , \nn \\
&V_{D6_2}=\frac{e^{\hat{K}/M_{Pl}^2}}{M_{Pl}^4}\left[\left|M_{Pl}^2\partial_{\phi^2}W_{D6_2}+(\phi^2)^*  \hat W \right|^2 + \left|M_{Pl}^2\partial_{\phi^3}W_{D6_2}+(\phi^3)^*  \hat W \right|^2\right. \nn \\
&\qquad \qquad \qquad \  \left.  + \left|M_{Pl}^2\partial_{\phi^1}W_{D6_2}-(\phi^1)^*  \hat W \right|^2 - 2 \, |\phi^1 \hat W| ^2 \right] \ , \nn \\
&V_{D6_3}=\frac{e^{\hat{K}/M_{Pl}^2}}{M_{Pl}^4}\left[\left|M_{Pl}^2\partial_{\phi^2}W_{D6_3}+(\phi^2)^*
\hat W \right|^2 +
\left|M_{Pl}^2\partial_{\phi^3}W_{D6_3}+(\phi^3)^*  \hat W
\right|^2 \right. \nn \\
&\qquad \qquad \qquad \  \left.  + \left|M_{Pl}^2\partial_{\phi^1}W_{D6_3}-(\phi^1)^*  \hat W \right|^2 - 2 \, |\phi^1 \hat W| ^2 \right] \ . \nn
\end{align}

We have summarized in Table \ref{softfinIIA} the pattern of soft
supersymmetry-breaking terms which results of plugging
(\ref{supervacIIA}), (\ref{finmud60}) and (\ref{finmud6i}) into
the above scalar potentials, for mixed supersymmetry breaking
vacua. Note that, unlike the case for quaternionic breaking, there
is at least one modulus that becomes massive for each type of
brane. This confirms the fact that this class of vacua is not
related by T-duality to the quaternionic one.

\TABLE{
\begin{tabular}{|c||c|c|c|c|}
\hline
& $D6_0$ & $D6_1$ & $D6_2$ & $D6_3$ \\
\hline
\hline
$\mu_{11}$ & $2e^{\frac{\hat K}{2}}(t_2f^{\hat 2}_{\hat 1 3}-t_3f^{\hat 3}_{\hat 1 2})$ & $2e^{\frac{\hat K}{2}}(t_2f^{2}_{\hat 3\hat 1}+t_3f^{3}_{\hat 1 \hat 2})$ & $0$ & $0$ \\
$\mu_{22}$ & $0$ & $0$ & $2e^{\frac{\hat K}{2}}(t_2f^{2}_{\hat 1 \hat 3}+t_3f^{3}_{\hat 1 \hat 2})$ & $2e^{\frac{\hat K}{2}}(t_2f^{\hat 2}_{3\hat 1}+t_3f^{\hat 3}_{2\hat 1})$ \\
$\mu_{33}$ & $0$ & $0$  & $2e^{\frac{\hat K}{2}}(t_2f^{\hat 2}_{\hat 1 3}+t_3f^{\hat 3}_{\hat 1 2})$ & $2e^{\frac{\hat K}{2}}(t_2f^{2}_{\hat 3\hat 1}-t_3f^{3}_{\hat 1 \hat 2})$ \\
\hline
$m_{1\bar 1}^2$ & $|\mu_{11}|^2$ & $|\mu_{11}|^2$ & $-|m_{3/2}|^2$ & $-|m_{3/2}|^2$\\
$m_{2\bar 2}^2$ & $0$ & $0$ & $|\mu_{22}|^2+|m_{3/2}|^2$ & $|\mu_{22}|^2+|m_{3/2}|^2$ \\
$m_{3\bar 3}^2$ & $0$ & $0$ & $|\mu_{33}|^2+|m_{3/2}|^2$ & $|\mu_{33}|^2+|m_{3/2}|^2$\\
\hline
$B_{11}$ & $0$ & $0$ & $0$ & $0$\\
$B_{22}$ & $0$ & $0$ & $2 \mu_{22}\, \bar m_{3/2}$ & $2 \mu_{22}\,\bar m_{3/2}$\\
$B_{33}$ & 0 & $0$ & $2 \mu_{33}\,\bar m_{3/2}$ & $2 \mu_{33}\,\bar m_{3/2}$ \\
\hline
$A_{123}$ & $0$ & $0$ & $g_{D6_2}m_{3/2}$ & $g_{D6_3}m_{3/2}$ \\
\hline
$C_{1\bar 2\bar 3}$ & $\mu_{11}\,g_{D6_0}$ & $\mu_{11}\,g_{D6_1}$ & $0$ & $0$\\
$C_{\bar 1 2\bar 3}$ & $0$ & $0$ & $\mu_{22}\,g_{D6_2} $ & $\mu_{22}\,g_{D6_3}$\\
$C_{\bar 1\bar 2 3}$ & $0$ & $0$ & $\mu_{33}\,g_{D6_2}$ & $\mu_{33}\,g_{D6_3}$ \\
\hline
\end{tabular}

\caption{Torsion induced soft parameters for $D6_M$-branes, in a
no-scale vacuum of a factorizable twisted torus with $W$
independent of $T_1, U_2, U_3$.  The gauge coupling constants are
$g_{D6_0}=(S+\bar S)^{-1/2}$ and $g_{D6_k}=(U^k+\bar U^k)^{-1/2}$, and we have set
$M_{Pl}=1$.} \label{softfinIIA} }

Assuming (\ref{HZ}) for $H_i$, and making use of (\ref{condmass})
and (\ref{KD66}), it is possible to check that
$\mu_{ij,\rm{GM}}=m^2_{i\bj,\rm{soft}}=0$ for all the scalars in
the worldvolume of the $D6_0$ and $D6_1$-branes. Hence, the
$\mu$-terms for these branes, shown in table \ref{softfinIIA},
correspond to purely supersymmetric (holomorphic in bulk moduli)
$\tilde \mu$-terms. Moreover, $\RE \mathcal{W}_1\sim m_{3/2}$
gives rise to soft couplings in the worldvolume of the $D6_2$ and
$D6_3$-branes which satisfy the relations (\ref{n4finito}). To
this regard, the induced soft masses for $\phi^1$ are always
tachyonic, signaling an instability of the $D6_2$ and
$D6_3$-branes at the origin, within this type of vacua. It is
tempting to identify this instability with a Higgs mechanism. The
final state, however, is not captured by the potentials
(\ref{d6gg}), as they were derived under the assumption $\langle
\phi^i\rangle = 0$. Analogous tachyonic masses were obtained
in heterotic compactifications with asymmetric K\"ahler
domination~\cite{tachyon}.\footnote{We are grateful to Luis Ib\'a\~nez for this observation.}
It would be desirable to obtain a better
understanding of the nature of these tachyonic modes within this context.

Finally, as an illustration of how the above equations apply in a concrete model,
consider the example of section (\ref{IIAnoscalemixedex}). The non trivial
structure constants can be read from (\ref{mixedstr}),
\begin{equation}
f^2_{\hat 3 \hat 1}=f^3_{\hat 1 \hat 2}=f^{\hat 2}_{3\hat 1}=f^{\hat 3}_{\hat 1 2}=1\ .
\end{equation}
From (\ref{d6gg}) then we obtain the following tree-level scalar potentials for
the D-brane moduli,
\begin{align}
V_{D6_0}&=\left|\frac{\phi^2\phi^3}{\sqrt{2s}}-\frac{\phi^1}{s(8t_1u_2u_3)^{\frac{1}{2}}}\right|^2\ , \\
V_{D6_1}&=\left|\frac{\phi^2\phi^3}{\sqrt{2s}}+\frac{\phi^1}{s(8t_1u_2u_3)^{\frac{1}{2}}}\right|^2\ ,\nn \\
V_{D6_2}&=\left|\frac{\phi^1\phi^2}{\sqrt{2u_2}}-\frac{(\phi^3)^*}{(8t_1u_2u_3)^{\frac{1}{2}}}\right|^2+\left|\frac{\phi^1\phi^3}{\sqrt{2u_2}}+\frac{(\phi^2)^*}{(8t_1u_2u_3)^{\frac{1}{2}}}\right|^2+\left|\frac{\phi^2\phi^3}{\sqrt{2u_2}}+\frac{(\phi^1)^*}{(8t_1u_2u_3)^{\frac{1}{2}}}\right|^2 - \frac{|\phi^1|^2}{4t_1 u_2 u_3},\nn \\
V_{D6_3}&=\left|\frac{\phi^1\phi^2}{\sqrt{2u_3}}-\frac{(\phi^3)^*}{(8t_1u_2u_3)^{\frac{1}{2}}}\right|^2+\left|\frac{\phi^1\phi^3}{\sqrt{2u_3}}+\frac{(\phi^2)^*}{(8t_1u_2u_3)^{\frac{1}{2}}}\right|^2+\left|\frac{\phi^2\phi^3}{\sqrt{2u_3}}+\frac{(\phi^1)^*}{(8t_1u_2u_3)^{\frac{1}{2}}}\right|^2-
\frac{|\phi^1|^2}{4t_1 u_2 u_3}\ .\nn
\end{align}

\section{Conclusions} \label{concl}

We have explored the conditions to have no-scale supergravity
vacua on orientifolds of SU(3) structure manifolds, with
supersymmetry spontaneously broken at tree-level. Although we have
covered a broad set of supergravity backgrounds, we have found
only two classes of solutions, depending on whether the
supersymmetry breaking is mediated by neutral matter descending
from $\mathcal{N}=2$ hypermultiplets or from a mixture of vector
and hypermultiplets. The first case, which we have denoted
``quaternionic breaking'', corresponds to T-duals of the known
type IIB no-scale vacua with 3-form fluxes, and is fully
characterized by a single ISD poly-form mixing fluxes and
torsion. The second, labelled as ``mixed breaking'', is instead
related to fluxless Scherk-Schwarz compactifications and can be
characterized by a purely imaginary poly-form.

We have also computed the effective $\mu$-terms induced by the
torsion of the SU(3) structure manifold in the gauge theory of D5, D6 and
D9-branes, for vacua based on twisted tori. These encode the
tree-level dynamics of the branes in the supergravity vacuum. The
resulting patterns for type IIB (IIA) vacua, summarized in tables
\ref{mususy} and \ref{mususyIIA}, can be nicely arranged in terms
of the holomorphic (symplectic) properties of the structure
constants. A similar fact was already observed in
\cite{hepth0408036,0501139} for the D7-brane flux induced
$\mu$-term. The present patterns, however, contain a much richer
structure, allowing for mass terms for mostly all the brane
moduli. The potential applications for model building are
therefore promising.

Notice that, due to the presence of flat directions, every attempt
of extracting phenomenological information from these vacua should
also take the quantum dynamics into account. In this sense, the
patterns of soft terms for pure moduli mediation presented in
section \ref{soft} are partial and, in a concrete phenomenological
model, should be completed with non-perturbative and loop
contributions.

Since a full string theory treatment of non-perturbative effects
is missing, one is usually advocated to implement those at the
level of the effective field theory. From this perspective, the
structure of $\mu$-terms turns out to be also determinant, as the
non-perturbative dynamics is constrained by the number of
fermionic zero modes.

No-scale solutions of ten dimensional supergravity have been
considered very frequently in the framework of type IIB Calabi-Yau
orientifolds with O3/O7-planes and 3-form fluxes. In this case,
the supersymmetry is often restored when non-perturbative effects
are present. To this regard, we expect a similar behavior for the
full no-scale quaternionic breaking family of vacua. It would be
very interesting however to extend this analysis to the case of
mixed breaking studied here, and to check in particular if the
breaking of supersymmetry is actually propagated to the complete
solution. It would also be nice to understand the tachyonic
instability observed for one of the brane moduli in this family of no-scale vacua.

Finally, there are also other directions which we believe may
deserve further research. The conditions for supersymmetric vacua
allow for more general structures, such as $SU(3)\times SU(3)$. It is natural to
expect that these solutions also admit non-supersymmetric marginal
deformations analogous to the ones discussed here. It may be
interesting to look for new families of no-scale vacua within this
context. Understanding the structure of the effective supergravity
is a major task for phenomenological applications of string
theory. We hope to come back soon to these issues.

\acknowledgments

We thank E. Dudas, A. Font, T. Grimm, L. Ib\'a\~nez, H. Jockers,
A. Tomasiello, A. Uranga and D. Waldram for useful discussions.
The work of P.G.C. is partially supported by INTAS grant,
03-51-6346, RTN contracts MRTN-CT-2004-005104 and
MRTN-CT-2004-503369, CNRS PICS \#~2530, 3059 and 3747,
 and by a European Union Excellence Grant,
MEXT-CT-2003-509661. The work of M.G. is supported in part by ANR
grant BLAN06-3-137168. Additional support comes from RTN
contract MRTN-CT-2004-512194.

\vspace{0.5cm}

\appendix

\section{Conventions and spinors} \label{conv}

We take orientation conventions for which
\begin{equation}
*J=\frac{1}{2}J\wedge J\ , \quad \int J\wedge J\wedge J > 0\ .
\end{equation}
$J$ and $\Omega$ can be obtained from the SU(3) invariant spinor
$\eta$ and the metric by
\begin{align} \label{bilinears}
&\eta_{+}^{\dagger}\gamma^m \eta_{\pm}=0 \ , &  &{\eta}^{\dagger}_{-}
\gamma^{mnp} \eta_{+} =
      \tfrac{1}{2}i \, \left(\frac{\mathcal{N}_J}{\mathcal{N}_{\Omega}}\right)^{\frac{1}{2}}\Omega^{mnp} \ , \\
&{\eta}^{\dagger}_{\pm} \gamma^{mn}\eta_{\pm} =
      \pm \tfrac{1}{2}i \, J^{mn} \ , &  & {\eta}^{\dagger}_{+} \gamma^{mnp} \eta_{-} =
      \tfrac{1}{2}i \, \left(\frac{\mathcal{N}_J}{\mathcal{N}_{\Omega}}\right)^{\frac{1}{2}} \bar {\Omega}^{mnp}\ , \nn
\end{align}
where $\mathcal{N}_{J}$ and $\mathcal{N}_{\Omega}$ are given in
(\ref{NJNO}) and $\eta^\dagger_\pm \eta_\pm = \tfrac12$,
$\eta_+^*=\eta_-$ (i.e. we are using the intertwiner between
$\gamma_m$ and $-\gamma_m^*$ to be 1).

The Mukai pairing between forms is related to the norm of
bispinors by \cite{GLW2}, \beq \label{mukaispi} \int \langle \Phi, \chi
\rangle = \frac{1}{2} \textrm{tr} (i \gamma_{7} \Phi^T_{\epsilon}
\chi_{\epsilon} ) {\cal N}_J\ , \eeq where $ \Phi_{\epsilon},
\chi_{\epsilon}$ are the bispinors corresponding to the forms
$\Phi, \chi$. We find
also convenient to use the Fierz identity \beq \label{fierz}
\eta_{+}{\tilde \eta}^\dagger_\pm =
      \frac{1}{4} \sum_{k=0}^6 \frac{1}{k!}
      \left(\tilde \eta^\dagger_\pm\gamma_{m_1\dots m_k}\eta_+\right)
         \gamma^{m_k\dots m_1} \ ,
\eeq
to write the forms in (\ref{SU3}).

\section{Decomposition in SU(3) representations} \label{su3decomp}

Part of the underlying approach that we use in the paper relies on
the decomposition of forms in $SU(3)$ representations. For
poly-forms, it is more convenient to use the generalized Hodge
diamond \cite{gualtieri,gualtieri2,GMPT1}, whose elements are
given by the different poly-forms in (\ref{Gplus}). Each component
is then computed by an appropriate integral. Concretely, the
different components of the 3-form decomposition (\ref{su3g3}) are
obtained from,
\begin{equation}
G_{(1)}^+=-\frac{i}{12\mathcal{N}_J}\int \Omega\wedge G\ , \quad
G_{(1)}^-=\frac{i}{12\mathcal{N}_J}\int \bar \Omega\wedge G \ ,
\quad G_{(3)}^{\pm}=\frac{1}{2} J\llcorner G^{\pm}\ .
\end{equation}
Analogously, the components of the even form $G$ in the $SU(3)$ decomposition
(\ref{Gplus}) are expressed in terms of the following integrals,
\begin{align} G_{(1)}^\pm&= \pm\frac{i}{8 {\cal
N}_{\Omega}} \int \langle e^{\mp iJ}, G \rangle \ , &
G^\pm_{mn}&=\pm\frac{i}{32 {\cal N}_J} J_{mp} J_{nq} \int \langle
\gamma^p e^{\pm iJ} \gamma^q, G
\rangle \ , \nn \\
G^+_m&= -\frac{1}{16{\cal N}_\Omega}J_{mn}\int \langle
\gamma^n\Omega, G\rangle \ , & \tilde{G}^+_m&= \frac{1}{16{\cal
N}_\Omega}J_{mn}\int \langle \overline{\Omega}\gamma^n,
G\rangle \ ,\nn \\
G^-_m&= -\frac{1}{16{\cal N}_\Omega}J_{mn}\int \langle
\gamma^n\overline{\Omega}, G\rangle \ , & \tilde{G}^-_m&=
\frac{1}{16{\cal N}_\Omega}J_{mn}\int \langle \Omega\gamma^n,
G\rangle \ , \end{align} with $\gamma^m$ given in (\ref{gammas}).

For real single-degree even forms, we use also the following SU(3)
decomposition, \bea F_{2}&=& \frac{1}{3} F_{2}^{(1)}\, J +  \RE
(F_{2}^{(3)}\llcorner
\overline{\Ox}) + F_{2}^{(8)}  \, , \nn \\
F_{4}&=& \frac16 F_{4}^{(1)} J \wedge J +   \RE (F_{4}^{(3)}\wedge
\overline{\Ox}) + F_{4}^{(8)} \, , \nn\\
F_{6} &=& \frac{1}{6}  F_{6}^{(1)} \,J \wedge J \wedge J \, . \eea
These singlets are a combination of the four singlets
$G^\pm_{(1)}$, $G^\pm_{mn}J^{mn}$ defined in  (\ref{Gplus}).

Finally, for $F_3$ and $H$, we use
\begin{align}
F_3 &= \frac{3\mathcal{N}_J}{\mathcal{N}_{\Omega}} {\rm Re}(F_{(1)} \bar{\Omega}) + F_{(3)} \wedge J
+ F_{(6)} \ , \label{f3desc}\\
H &= \frac{3\mathcal{N}_J}{\mathcal{N}_{\Omega}} {\rm Re}(H_{(1)} \bar{\Omega}) + H_{(3)}
\wedge J + H_{(6)} \, .
\end{align}
where comparing to (\ref{su3g3}), $F_{(1)}=F^+_{(1)}= (F^-_{(1)})^*$. In O6 compactifications $H$ is odd under the orientifold action, same as $\IM \Omega$.
This implies that $H^{(1)}$ is real.

\section{Torsion classes on twisted tori} \label{torsiontori}

For completeness in this appendix we present the torsion classes
for a twisted torus in terms of the structure constants
$f^a_{bc}$ defined in (\ref{torsion}). For alternative expressions,
the reader may also consult \cite{robbins1}.

Defining the spin connection 1-form with holomorphic indices, $\omega^{mn}$,
through,
\begin{equation}
dz^m+\omega^{mn}\wedge z_n + \omega^{m\bar{n}}\wedge
\bar{z}_n = 0\ ,
\end{equation}
with holomorphic vectors $z^m=e^m+iU^m{}_ne^n$, for $m=1,2,3$, and
acting with the exterior derivative on $\Omega$ and $J$ given in
(\ref{omtwist}), we extract the torsion classes,
\begin{align}
\mathcal{W}_1&=\frac{2i}{3}\epsilon_{mno}\imath_{z^m}\omega^{no}\ , \\
\mathcal{W}_2&=-\epsilon_{mno}\omega^{mn}\wedge z^o-\mathcal{W}_1J\ , \\
\mathcal{W}_3&=\frac{i}{2}\omega_{mn}\wedge z^m\wedge z^n+\frac{3}{4i}\frac{\mathcal{N}_J}{\mathcal{N}_{\Omega}}\overline{\mathcal{W}}_1\Omega\ + \ c.c.\ , \\
\mathcal{W}_4&=\mathcal{W}_5=0\ ,
\end{align}
with $\epsilon_{123}=-i$. In terms of (\ref{torsion})
the spin connection reads,
\begin{equation}
\omega^{ab}\equiv -\frac{1}{2}\left(\imath_{e^a}de^b-\imath_{e^b}de^a-e_c(\imath_{e^a}\imath_{e^b}de^c)\right)=
\frac{1}{2}(f^{b}_{cd}e^cg^{ad}-f^a_{cd}e^cg^{bd}-f^c_{de}g^{ad}g^{be}e_c)\ ,
\end{equation}
with $e_c\equiv g_{bc}e^b$. Hence, in terms of structure constants with holomorphic/antiholomorphic indices,
\begin{align}
\mathcal{W}_1&=\frac{i}{3}g^{m\bar r}g^{n\bar s}\epsilon_{mno}f^o_{\bar{r}\bar{s}}\ ,\label{w1twist}\\
\mathcal{W}_2&=-\mathcal{W}_1J+\epsilon_{mno}g^{n\bar s}\left(f^o_{\bar{p}\bar{s}}+\frac{g_{q\bar p}}{2}f^q_{\bar{o}\bar{s}}\right)z^m\wedge \bar{z}^p\ , \\
\mathcal{W}_3&=\frac{i}{2}\left(g_{m\bar s}f^{\bar{s}}_{n\bar{o}}-\frac{g_{r\bar o}}{2}f^r_{mn}\right)z^m\wedge z^n\wedge \bar{z}^o+c.c.\ , \\
\mathcal{W}_4&=\mathcal{W}_5=0\ .
\end{align}

\end{document}